\documentclass{svproc}

\usepackage{url}

\usepackage{graphicx}
\usepackage{xcolor}
\usepackage{pgfplots}

\usepackage{amsmath,amssymb,amsfonts}
\usepackage{textcomp}
\usepackage{booktabs, tabularx}
\usepackage{slashbox}
\usepackage{multirow}
\usepackage{float}
\usepackage[mathscr]{euscript}
\usepackage{subcaption}
\usepackage{cite}

\usepackage{hyperref}
\usepackage{cleveref}

\crefname{figure}{Fig.}{Figs.}
\crefname{table}{Table}{Tables}
\crefname{equation}{}{}

\begin{document}

\pgfplotsset{
    compat=1.7,
    grid=both,
    grid style={line width=1pt, draw=black},
    major grid style={line width=.3pt,draw=gray!30},
    tick align=inside,
    major tick length=2pt,
    tick label style={font=\tiny},
    axis line style=very thin,
    every axis/.append style={
        {font=\tiny},
    },
    /pgfplots/legend image code/.code={%
        \draw[mark repeat=2,mark phase=2] 
            plot coordinates {
                (0cm,0cm)
                (0.17cm,0cm)        
                (0.33cm,0cm)
            };
    },
    /pgfplots/x label style={at={(axis description cs:0.5,-0.13)},anchor=north},
    /pgfplots/y label style={at={(axis description cs:-0.13,0.5)},anchor=south},
    /pgfplots/legend style={
        line width=.3pt,
        draw=black,
        fill=white,
        at={(1.47,0.5)},
        anchor=east,
        legend columns=1,
        nodes={scale=0.85, transform shape},
        mark options={mark size=1.75pt},
        row sep=1pt},
    width=5.5cm,
    height=4.7cm,
}
\mainmatter              
\title{STAR: Spatio-Temporal Prediction \\ of Air Quality Using A Multimodal Approach}
\titlerunning{Spatio-Temporal Prediction of Air Quality Using A Multimodal Approach}  

\author{Anonymous}

\author{Tien-Cuong Bui\inst{1} \and Joonyoung Kim\inst{1} \and Taewoo Kang\inst{1} \and Donghyeon Lee\inst{1} \and Junyoung Choi\inst{1} \and Insoon Yang\inst{1} \and Kyomin Jung\inst{1} \and Sang Kyun Cha\inst{2} }

\institute{Dept. of Electrical and Computer Engineering, Seoul National University, Korea
\and
Graduate School of Data Science, Seoul National University, Korea 
\email{cuongbt91@snu.ac.kr}
}
\maketitle              
\begin{abstract}

With the increase of global economic activities and high energy demand, many countries have raised concerns about air pollution. However, air quality prediction is a challenging issue due to the complex interaction of many factors. In this paper, we propose a multimodal approach for spatio-temporal air quality prediction. Our model learns the multimodal fusion of critical factors to predict future air quality levels. Based on the analyses of data, we also assessed the impacts of critical factors on air quality prediction. We conducted experiments on two real-world air pollution datasets. For Seoul dataset, our method achieved 11\% and 8.2\% improvement of the mean absolute error in long-term predictions of PM\textsubscript{2.5} and PM\textsubscript{10}, respectively, compared to baselines. Our method also reduced the mean absolute error of PM\textsubscript{2.5} predictions by 20\% compared to the previous state-of-the-art results on China 1-year dataset.

\end{abstract}
\section{Introduction}
Air pollution is rapidly becoming a pressing issue for many large cities due to the rapid urbanization along with concentrated economic activities. These issues increase the need and demand for an accurate citywide air quality prediction model, which is vital for public health protection and government regulation. 

However, air quality (AQ) prediction is a challenging problem due to the involvement of multiple factors, including local pollutant emissions, coal power plants, dust activities, seasonal conditions, meteorology, terrain, and several other human activities \cite{kim2018characterization, kim2018trend}. Moreover, one of the main reasons for the difficulties of this problem is the lack of public datasets that include a large amount of data on critical factors. A straightforward approach is to consider AQ prediction as a typical time-series prediction problem \cite{bui2018deep}. Then, the predictive model repeats this process multiple times for each sequence of data. Since air quality depends on both temporal changes and spatial relations, several models have been proposed to predict air quality levels for monitoring stations \cite{zheng2013u, zheng2015forecasting}. However, the prediction performances of these methods depended on feature extraction processes, which are manually determined. Furthermore, most of the predictive models have focused on only hourly predictions, while long-term AQ level predictions have been overlooked.

In this paper, we propose a multimodal approach for spatio-temporal air quality prediction, entitled STAR. Our model focuses on predicting the two most concerned factors of air pollution PM\textsubscript{2.5} and PM\textsubscript{10} due to their chronic effects on human health, according to \cite{pope2002lung}.  It proceeds as follows: First, a spatial transformation function converts observational data (AQ data and meteorology) into heat-map images to analyze spatial changes of air quality. By using images, our model effectively captures spatial properties and relationships without explicitly modeling the spatio-temporal dynamics of PM\textsubscript{2.5} or PM\textsubscript{10} and other atmospheric factors. Secondly, the model combines the spatial representation with temporal features of critical factors into an encoded vector. This vector is then concatenated with the representation features of the output from the previous time step to form a context vector. Given this, our model generates a heat-map image with two feature maps for the prediction of particulate matters. 

To prove the efficacy of the proposed method, we perform experiments on real-world air pollution datasets. Experiments are first conducted on Seoul 5-year dataset, and then the model robustness is evaluated using China 1-year dataset \cite{zheng2015forecasting}. Specifically, we directly collect 10-year data (2008-2018), including observational meteorology and air pollutant concentrations from the Seoul Metropolitan Government's websites. Additionally, we attempt to collect air pollution data of some Chinese cities and weather forecasts of all cities to assess the impact of external factors on Seoul AQ levels. For Seoul dataset, we compare our model's results with baselines in hourly (up to 24 hours) and long-term (up to 7 days) predictions. Our method achieves 11\% and 8.2\% improvement of the mean absolute error in long-term predictions of PM\textsubscript{2.5} and PM\textsubscript{10} compared with other baselines on Seoul dataset. As the size of 1-year China dataset is small, it creates difficulties for training deep learning models. Hence, we apply a transfer learning method, which uses the pre-trained weights of Seoul dataset for China dataset, to overcome this issue. As a result, our method improves the mean absolute error of PM\textsubscript{2.5} predictions by 20\% compared to the previous state-of-the-art methods. We also train the model from scratch on this dataset, and the results are 11\% better than the previous method on the mean absolute error.

Based on real-world datasets, we assess the impact of critical factors on AQ prediction. We conduct various experiments on all combinations of data sources, including local AQ data, air pollution data of neighboring countries, and the meteorological data. For instance, the model is trained to predict future AQ levels using only meteorological data to assess the impact of weather conditions on AQ levels. The experimental results reveal that the local PM\textsubscript{2.5} levels are directly connected with the meteorological conditions and the PM\textsubscript{2.5} variations of neighboring countries, while PM\textsubscript{10} levels are localized and less sensitive to these factors. 

The remainder of this paper is structured as follows. In \cref{related_work}, we introduce the related work. In \cref{data_collection}, we describe datasets and collection methods. Then the structure and technical details of our proposed method are demonstrated in \cref{approach}. In \cref{experiment_sec}, experiments are presented. Finally, we conclude our work in \cref{conclusion_part}.

\section{Related Work} \label{related_work}
\subsection{Spatio-Temporal Prediction}
The spatio-temporal prediction has been prevalent in many domains, especially urban problems. Donahue et al., 2015 \cite{donahue2015long} proposed a class of architectures, which combined convolutional layers and long-range temporal (CNN-LSTM), for large-scale visual understanding tasks. Our work is similar to LRCNs in terms of the order of learning processes, in which it first models spatial dependencies through CNNs, then uses LSTM networks for capturing temporal dynamics. In urban problems, several works have applied graph convolutional networks (GCN) and RNNs for spatio-temporal prediction. For instance, Li et al., 2017 \cite{li2017diffusion} introduced a model to tackle time-series prediction problems in the traffic domain. Correspondingly, Yu et al., 2017 \cite{yu2017spatio} modeled the traffic flow as a diffusion process on a directed graph and applied a derived version of GCN to the forecasting process. In the traffic domain, Yao et al., 2018  \cite{yao2018deep} proposed a CNN-LSTM model for taxi demand prediction, which considered multiple features such as meteorology or events as dominant factors. He and his co-workers also proposed a spatio-temporal dynamic network to predict traffic volume \cite{yao2019revisiting}. 

In the AQ prediction domain, not many methods applied conventional CNN architecture due to the spatial sparsity of monitoring data. Although some papers used derived versions of CNNs for making predictions \cite{qi2019hybrid, lin2018exploiting}, they required exact geo-location information to specify the edges in the computation graph. On the contrary, we proposed an efficient way of transforming observational data into heat-map images and applied CNN-LSTMs to predict future AQ levels. 

In the time-series prediction domain, several works have been proposed to improve recurrent neural networks for specific problems. For instance, Qin et al., 2017 \cite{qin2017dual} introduced a dual-stage attention-based method to capture long-term dependencies. Correspondingly, Seo et al., 2018 \cite{seo2018structured} proposed a generalization of classical RNNs for predicting structured sequences of data.

\subsection{Air Quality Prediction}
This problem has received considerable attention from numerous researchers across diverse subjects. In the area of environmental science, researchers tried to identify the root cause of air pollution and the correlation with critical factors. These predictive algorithms based on classical methods \cite{zhu2016gaussian, zhu2015granger, drucker1997support} were challenging to apply in large-scale. Our method instead employed the methodology of multimodal fusion of numerous data sources to predict future air quality levels. Multimodal fusion is becoming increasingly prevalent as it proves to be a practical approach to learn features from various datasets for making accurate predictions. The advent of deep learning and big data facilitates the ubiquity of this approach. Zheng et al. proposed \cite{zheng2013u, zheng2015forecasting}, which were the very first research employing big data techniques to predict future AQ levels. However, these methods were based on fundamental machine learning methods, which required manual feature extraction and modeling. Lately, deep learning models for the spatio-temporal prediction focused on the variation of AQ levels in each monitoring station based on weighted networks \cite{cheng2018neural, yi2018deep}. These works are similar to us in feature embedding methods for meteorological conditions, time encoding, and AQ data. However, they did not assess the exact influence of critical factors on short-to-long-term AQ levels. Our approach, which integrates the spatio-temporal features of multiple data sources, addresses this problem and predicts short-to-long-term AQ levels. Similar to us, Le et al., 2019 \cite{le2019spatiotemporal} transformed observational data into heat-map images and utilized ConvLSTM \cite{convlstm} to handle spatiotemporal variations.

\section{Data Collection} \label{data_collection}
\noindent\textbf{Seoul 5-year Dataset}. It consists of multiple data sources, which were directly gathered from different public websites. 
\begin{itemize}
    \item \textbf{Air Pollution Data of Neighboring Areas}. We considered the air pollution sources of some cities in China (Beijing, Shandong, Shenyang) as an external factor, which affects Seoul AQ levels. PM\textsubscript{2.5} data of these cities from 2014 were collected from the Berkeley Earth website. All data were shifted one hour (+1h) to match the time zone of Seoul (Seoul GMT +9).
    \item \textbf{Seoul Air Pollution Data}. These data were collected from a website\footnote{http://cleanair.seoul.go.kr} of the Seoul Metropolitan Government. They include hourly observations of gaseous pollutants (NO\textsubscript{2}, SO\textsubscript{2}, CO\textsubscript{2}, O\textsubscript{3}) and particulate matters (PM\textsubscript{2.5}, PM\textsubscript{10}). Even though collected data were available from 2008, we utilized only five years from 2014 to 2018 for training, testing, and validating to coincide with the data of neighboring areas. The period 2014 to 2016 was utilized as the training set. Correspondingly, 2017 data and 2018 data were used as the validation set and test set. 
    \item \textbf{Meteorological Data}. We considered meteorology as a determining factor of future AQ levels. First, we crawled weather forecast data of all cities (Seoul, Beijing, Shenyang, and Shandong) from World Weather Online. Second, we obtained meteorological observational data of all areas in Seoul from the government's system\footnote{Seoul Autonomous Weather System}. We aggregated the observational data, which have been recorded every minute since 2008, to one-hour intervals. Both data sources include standard meteorological information such as wind speed, wind direction, humidity, temperature, and precipitation.
    \item \textbf{Additional Information}. To represent the impact of other factors, for which data were either unavailable or difficult to obtain, we added to input vectors categorical features such as months in a year, hours in a day, and a binary holiday flag. These features can indirectly represent additional information such as seasonal variations, transportation situation, to name a few. For holiday information, we collected the data for all cities from the Time and Date website. 
\end{itemize}

\noindent\textbf{China 1-year dataset}. This dataset was published in Zheng et al., 2015 \cite{zheng2015forecasting}. It consists of AQ data, meteorological data, and weather forecasts of many Chinese cities. However, its small volume and a large amount of missing data cause difficulties for deep learning approaches. 

\section{Proposed Method} \label{approach}
\subsection{Notation and Problem Statement}
First, we define data sources used in our model as follows:
\begin{itemize}
    \item $\mathcal{I} = \big\{I^k_t\big\}$: AQ data of an \textbf{I}nvestigated area
    \item $\mathcal{M} = \big\{M^k_t, W^x_t\big\}$: \textbf{M}eteorological data 
    \item $\mathcal{N} = \big\{N^x_t\big\}$: AQ data of \textbf{N}eighboring countries       
    \item $\mathcal{D} = \big\{D^x_t\big\}$: Additional \textbf{D}ate information
\end{itemize}

\setlength{\tabcolsep}{0.85em}
\begin{table}[b!] 
\caption{Input Vector Features. They consist of categorical and numeric data elements of four data sources.}
\begin{center}
\begin{tabular}{cccc}
\hline
Data Source & Feature & Range & Type \\
\hline
\multirow{6}{*}{$\mathcal{I}$}       & PM\textsubscript{2.5} & $[0,1]$ & \multirow{6}{*}{Float} \\ 
                          & PM\textsubscript{10} & $[0,1]$ & \\ 
                          & O\textsubscript{3} & $[0,1]$ & \\ 
                          & NO\textsubscript{2} & $[0,1]$ &  \\ 
                          & SO\textsubscript{2} & $[0,1]$ &  \\ 
                          & CO & $[0,1]$ & \\
\hline
\multirow{6}{*}{$\mathcal{M}$}    & Temperature & $[0,1]$ & \multirow{6}{*}{Float} \\ 
                                & Humidity & $[0,1]$ &  \\ 
                                & Precipitation & $[0,1]$ &  \\ 
                                & Wind Speed & $[0,1]$&  \\ 
                                & Wind Gust & $[0,1]$ &  \\  
                                & Wind Direction & $[0,1]$ & \\
\hline
$\mathcal{N}$ & PM\textsubscript{2.5} & $[0, 1]$ & Float \\
\hline
\multirow{3}{*}{$\mathcal{D}$}           & Month & $[1,12]$ & \multirow{3}{*}{Categorical} \\ 
                                & IsHoliday & $\{0,1\}$ &  \\ 
                                & Hour & $[1,24]$ &  \\ 
\hline
\end{tabular}
\end{center}
\label{embedding_tbl}
\vspace{-5mm}
\end{table}

For brevity, index $t$, $k$, and $x$ represent each time step, each station, and each city, respectively. Specifically, $x$ can be either $s$ or $c$, which represent an investigated area or cities from neighboring countries, respectively. $I^k_t$ is a 6-dimensional vector of AQI concentrations in station $k$ at time-step $t$. Similarly, $N^x_t$ denotes a 1-dimensional vector of PM\textsubscript{2.5} concentrations of a city $x$ for time step $t$. $M^k_t$ and $W^x_t$ mean a 6-dimensional vector of meteorological conditions, including temperatures, humidity, precipitation, wind, and weather forecasts. Additionally, $M^k_t$ denotes meteorological observational data of a station $k$ at time-step $t$, while $W^x_t$ is weather forecasts of a city $x$. Lastly, $D^x_t$ denotes a 3-dimensional vector of date information of a city $x$, including a binary holiday flag, months, and hours.

Based on the statistical data analysis, we specified the input features of data sources, as described in \cref{embedding_tbl}. The features of $\mathcal{I}$ and $\mathcal{M}$ consist of all AQI and standard meteorological conditions. Unlike $\mathcal{I}$, $\mathcal{N}$ has only PM\textsubscript{2.5} due to the lack of open-source data for cities from neighboring countries. Besides, $\mathcal{D}$'s features are different, corresponding to each country.\\

\noindent\textbf{Problem Statement} Given $\mathcal{I, M, N,D}$, the model $f_{\theta}$ aims to predict $\widehat{Y}$, which is the AQ levels (PM\textsubscript{2.5} and PM\textsubscript{10} concentrations) over the next $T_{pred}$ hours for each station $k \in K$.
\begin{equation}
    f_{\theta}\Big( \mathcal{I, M, N, D}\Big) = \widehat{Y}
\end{equation}
Our goal is to train the model $f_{\theta}$ using collected datasets with $L_2$ loss functions.

\subsection{Model} \label{model_lbl}
\begin{figure*}[t]
\centering
\includegraphics[trim={2.72cm 3.1cm 2.97cm 1.67cm},clip,width=\linewidth]{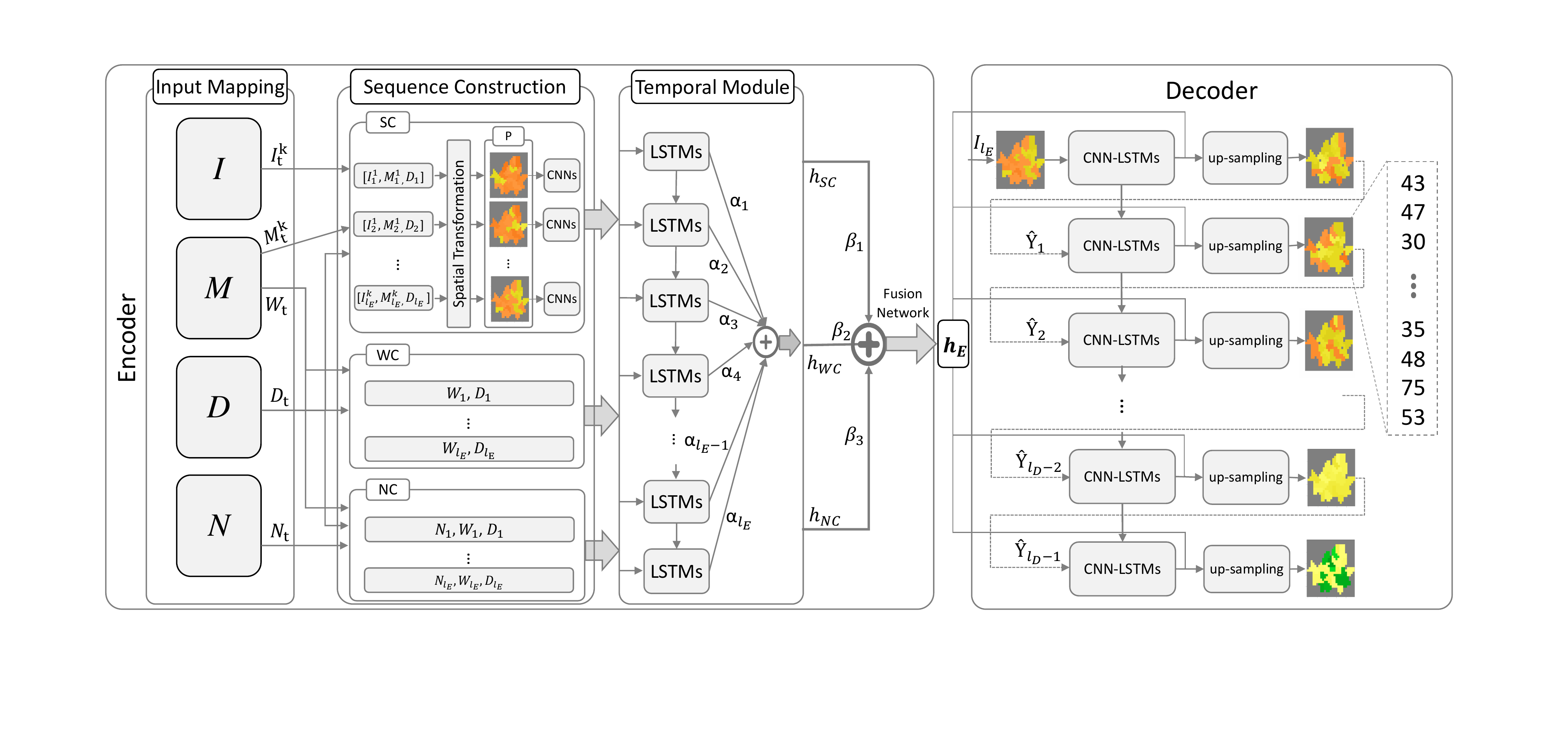}
\caption{An overview of STAR. It is based on an encoder-decoder framework with CNN-LSTM networks. The encoder consists of three modules, which process input vectors of four data sources and output a 128-dimensional vector $h_E$. The decoder generates heat-map images of future air quality based on CNN-LSTM units and up-sampling networks. Notably, $l_E$ and $l_D$ are arbitrary parameters, which can be varied in different modules. The weights $\alpha$ and $\beta$ are computed by a softmax function.}
\label{apnet_overview}
\end{figure*}

To capture features from various data sources, we designed a multimodal approach based on an encoder-decoder framework. As depicted in \cref{apnet_overview}, our model comprises two main components an encoder and a decoder. 
\subsubsection{Encoder}
The encoder comprises an input mapping module, a sequence construction module, a temporal module, and a fusion network. \\
\noindent\textit{Input Mapping}. Based on domain knowledge, we discern that direct and indirect factors have different effects on AQ levels. Therefore, we concatenate features of four data sources $\mathcal{I}, \mathcal{M}, \mathcal{N}, \mathcal{D}$ into vectors $\mathcal{SC}$, $\mathcal{WC}$, and $\mathcal{NC}$, as shown in \cref{eq:mapping}. These vectors are the input of three subnets \textbf{SC}, \textbf{WC}, and \textbf{NC} in the sequence construction module.

\begin{align}
    \mathcal{SC} &= \text{Concatenate}\big(\mathcal{I}, \mathcal{M},\mathcal{D}\big) \nonumber \\ 
    \mathcal{WC} &= \text{Concatenate}\big(\mathcal{M}(W^s),
    \mathcal{D}\big) \label{eq:mapping}\\ 
    \mathcal{NC} &= \text{Concatenate}\big(\mathcal{N}, \mathcal{M}(W^c),\mathcal{D}\big) \nonumber
\end{align} 

\noindent\textit{Sequence Construction}. The sequence construction module comprises three parts \textbf{SC}, \textbf{WC}, and \textbf{NC}, which aim to construct sequences of input vectors from data sources for the temporal module. The reason for building such three subnets is the effect of different factors on air quality. Furthermore, it is infeasible to use only one network to extract meaningful information from numerous features. \textbf{SC} captures the spatial relations in observational data, while \textbf{WC} and \textbf{NC} represent temporal changes in future meteorology (weather forecasts) and air pollution sources from neighboring countries. The roles of each component can be elaborated as follows:

\begin{figure}[t!]
\centering
\includegraphics[trim=5.98cm 4.06cm 7.69cm  2.0cm,clip,width=0.96\linewidth]{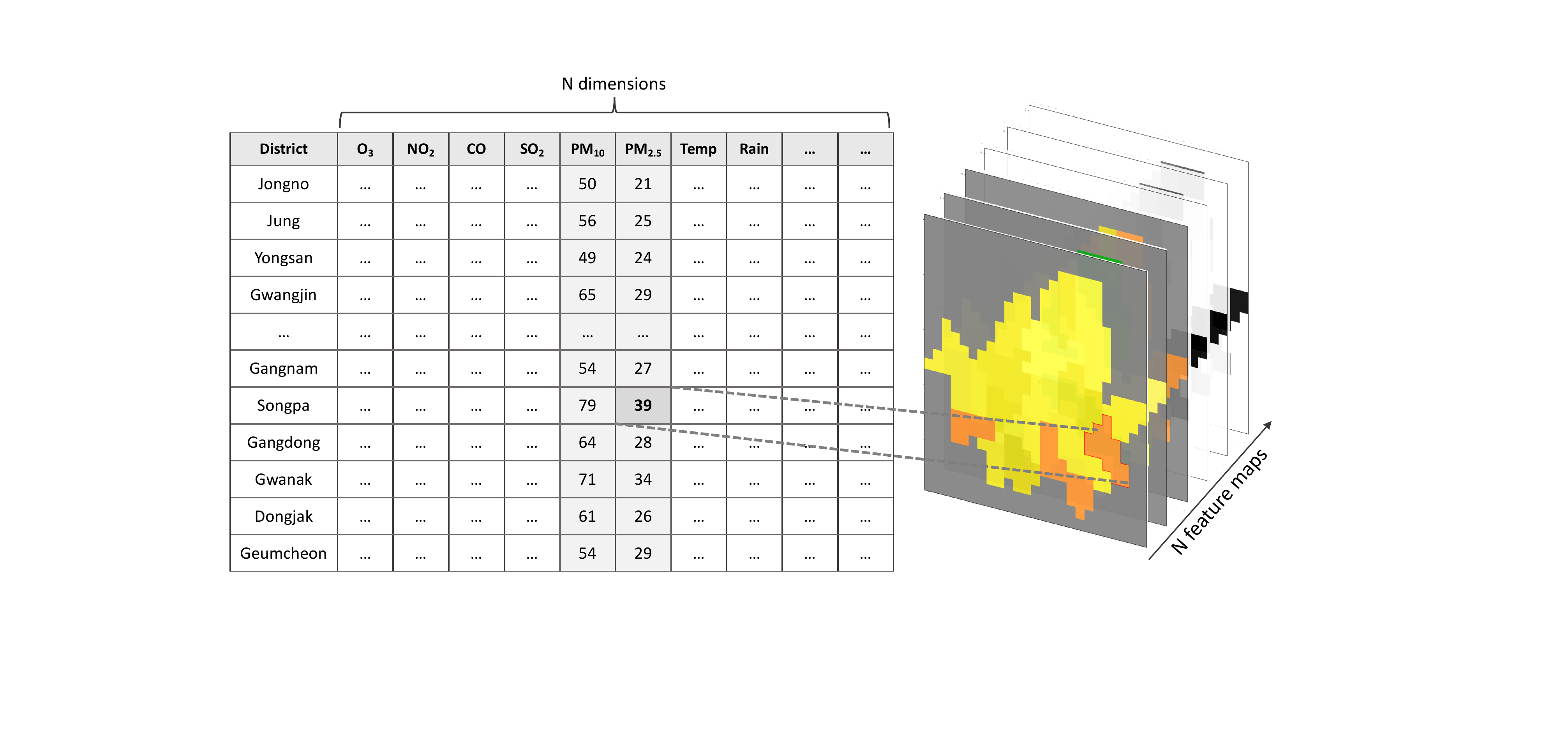}
\caption{Spatial transformation function. It transforms tabular data into heat-map images. Each area's pixels hold the same value as its observational data.}
\centering
\label{row_space}
\end{figure}

\begin{itemize}
    \item \textbf{SC}'s inputs are the combination of observational data and additional date information of the investigated area. Each vector comprises of 15 features of $\mathcal{I}$, $\mathcal{M}$, and $\mathcal{D}$ in \cref{embedding_tbl}. \textbf{SC} has a spatial transformation function, which turns input vectors into heat-map images to capture spatial relations in the observational data among areas. Finally, it generates a sequence of 128-dimensional vectors.
    \item \textbf{WC} captures the temporal changes in future weather conditions of investigated areas. \textbf{WC}'s inputs are 9-dimensional vectors, which comprise of features of $\mathcal{M}$ and $\mathcal{D}$.
    \item \textbf{NC} is responsible for determining the impact of transboundary air pollution sources on local AQ levels. Its inputs are 24-dimensional vectors, including 21 (7x3) features of $\mathcal{N}$ and $\mathcal{M}$ of three Chinese cities, and 3 features of $\mathcal{D}$. These three features are critical to revealing AQ variations of neighboring areas since $\mathcal{N}$ has only PM\textsubscript{2.5} data.
\end{itemize}

\begin{figure}[t]
    \begin{subfigure}[t]{0.50\linewidth}
        \centering
        \includegraphics[trim={7.66cm 7.24cm 25.54cm 2.24cm},clip,width=3cm]{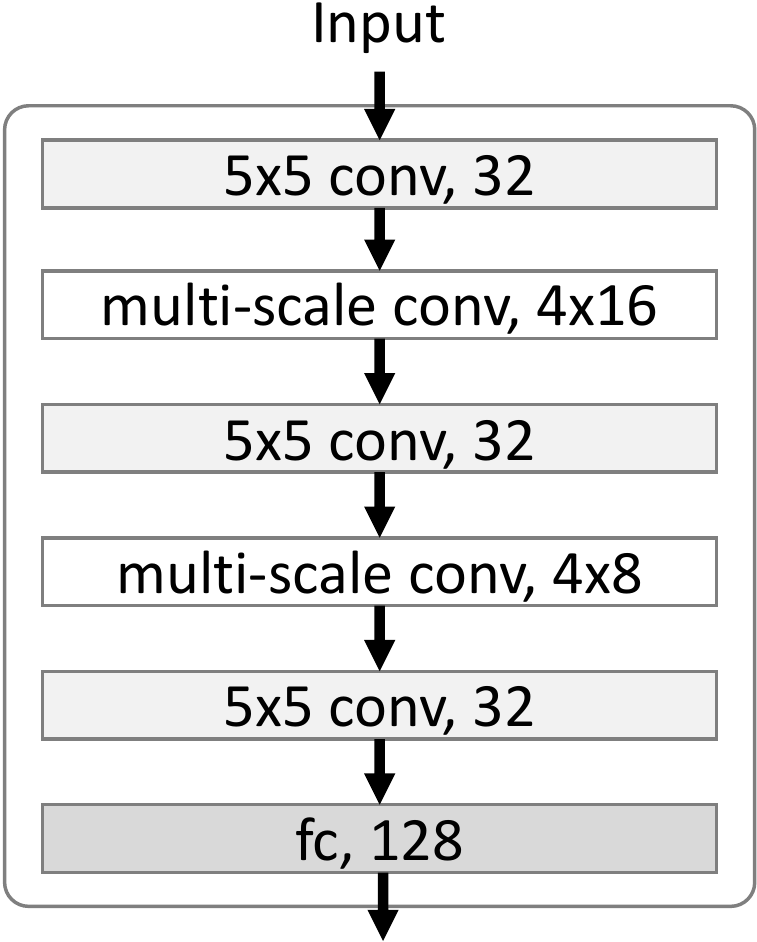}
        \caption{CNN layers}
        \label{mscnn}
    \end{subfigure}
    \begin{subfigure}[t]{0.45\linewidth}
        \centering
        \includegraphics[trim={19.84cm 6.6cm 13.09cm 4.13cm},clip,width=3.5cm]{images/cnn_structure.pdf}
        \caption{Up-sampling unit}
        \label{trans_cnn}
    \end{subfigure}
    \caption{(a) depicts the convolutional layers used in the \textbf{SC} module and the CNN part of a decoder's CNN-LSTM unit. (b) shows the structure of the Up-sampling unit used for generating prediction outputs. }
    \label{tblapnet}
\end{figure}

\noindent\textit{Spatial Transformation}.
This function is implemented inside \textbf{SC} module to convert tabular data into heat-map images of 25 $\times$ 25 size as depicted in \cref{row_space}. As the observational data of investigated areas are organized as districts, heat-map transformation allows STAR to effectively capture spatial interactions without explicitly modeling the spatial-temporal dynamics of various factors.  We develop a tool to determine the boundary of districts in a 25 $\times$ 25 grid. Then, a transformation function fills all pixels of a district with a value corresponding to a dimension of an input vector. This process is repeated many times for all dimensions. \cref{apnet_overview} depicts a group of heat-map images as \textbf{P}. Next, our model pushes these images to a convolutional-based network to capture spatial relations. A multi-scale convolutional neural network (MS-CNN) is modeled as similar to the naive version of \cite{szegedy2015going} to enable our model to learn both local and global spatial features of the images. It consists of four kernel sizes $[1,3,5,7]$. \cref{mscnn} demonstrates the detailed structure of the convolutional layers used in the spatial transformation function.

\noindent\textit{Temporal module}. The temporal module, which uses attention-based networks with LSTM units, is responsible for outputting a 128-dimensional hidden vector from a sequence of vectors. Our model employs LSTM to learn the temporal changes of critical factors. A weighted attention layer is attached on top of outputs to let our model learn the varying importance of each time step. 

\noindent\textit{Fusion Network}. Lastly, a fusion network combines hidden vectors of different components into a context output $h_E$ by using an attention layer. We use the attention network instead of a fully-connected layer since each factor has different impacts on AQ levels. The output vector is a 128-dimensional vector.

\subsubsection{Decoder}
The decoder consists of CNN-LSTM networks and an up-sampling unit. First, CNN-LSTM networks, which consist of convolutional layers and an LSTM cell, are responsible for capturing spatio-temporal features of prediction outputs. The convolution layers are the same as the ones used in \textbf{SC} module, as depicted in \cref{mscnn}. Second, the up-sampling unit aims at generating heat-map images of future AQ levels. It consists of three transposed-convolutional layers, with a filter size of $5\times5$, as shown in \cref{trans_cnn}. At each prediction time step $t$, a 256-dimensional vector is generated by concatenating the CNN-LSTM output $\widehat{y}_{t-1}$ with $h_E$. This concatenated vector is reshaped to $2\times2\times64$ and fed to the up-sampling unit. The output of prediction time step $t$ becomes the input of the next time step $t+1$, except the first time step, whose input is the last input image of the encoder. The prediction output is an image with two feature maps corresponding to PM\textsubscript{2.5} and PM\textsubscript{10}. In testing, our model averages all pixels of each district in a heat-map image into a prediction value. \\

\section{Experiments} \label{experiment_sec}
\subsection{Baselines}
Here, we briefly explain nine alternatives for comparison as follows: 
\begin{itemize}
    \item \textbf{ARIMA}: it is a generalization of an autoregressive moving average model for predicting the next steps. We do not report the experimental results on daily predictions since its errors drastically increase.   
    \item \textbf{SVR}: Support Vector Regression is a variant of SVM for time-series prediction. The epsilon value of the radial basis function kernel is set as $0.1$. 
    \item \textbf{CNN}: it constitutes of three 1-D convolution layers with kernel size 5, $[256,128,64]$ filters, and pooling layers. We only report CNN's prediction results with $\mathcal{I}$'s features, since that is its best performance.
    \item \textbf{RNN}: it consists of $[64,64]$ encode-decoder states with LSTM units. Similar to CNN, we also use only $\mathcal{I}$'s features as the input of the model.
    \item \textbf{DA-RNN} \cite{qin2017dual}: DA-RNN is a dual-stage attention-based recurrent model for time-series prediction. We use all features as the input of the model.
    \item \textbf{GCRN} \cite{seo2018structured}: GCRN fuses CNNs and RNNs for predicting time-varying graph-based data. The localized filter is set as 2. We apply GCRN directly to the observation data of investigated areas, including meteorological and atmospheric features. Then we concatenate the output vector with other temporal outputs to predict future AQ levels.
    \item \textbf{DCRNN} \cite{li2017diffusion}: DCRNN uses bidirectional graph random walk and RNNs to capture spatio-temporal dynamics. The diffusion step value is set as 2. We implement the model as similar to GCRN.
    \item \textbf{ConvLSTM} \cite{le2019spatiotemporal, convlstm}: it is an extended version of FC-LSTM to handle spatiotemporal data. We use two convolution layers with a kernel size of $3\times3$. The number of filters is set as 8. Its encoder module is similar to STAR, while the decoder is replaced with fully connected layers.
    \item \textbf{DeepAir} \cite{yi2018deep}: it is a DNN-based approach, which consists of a spatial transformation module and a fusion network, and similar to us in the embedding method for critical factors.
\end{itemize}

\subsection{Experimental Settings}
\subsubsection{Implementation Details}
In this part, we describe hyper-parameters and configurations of experiments. 

\noindent\textit{Pre-processing}: We used min-max normalization to normalize features into $[0,1]$. In several works such as \cite{yi2018deep}, categorical features were normalized using one-hot encoding. However, we observed that one-hot encoding drastically increased the number of dimensions, thus resulting in the decline of prediction performance. Practically, we also transformed \textit{Month} and \textit{Hour} into the $[0,1]$ range by dividing them by 12 and 24, respectively.

\noindent\textit{Hyper-parameters}: We set the size of fully-connected layers as 128 dimensions and used \textit{Tanh} as the activation function. All weights were initialized by the Xavier initialization. The sequence length of \textbf{SC} and \textbf{NC} was defined as 24 and 48, while the length of \textbf{WC} was the same as $T_{pred}$. 

\noindent\textit{Optimization Method}: The ADAM optimizer was used with \(\beta_1 = 0.5\) and \( \beta_2 = 0.999\) for training, the learning rate was defined as 2x10\textsuperscript{-3}, and the batch size was 64. The dropout rate was set as 0.5 for multiple layers, and the early stopping technique was utilized in training to avoid overfitting. Our model used the $L_2$ loss function, as shown in \cref{l2_loss}.

\begin{equation} \label{l2_loss}
\mathcal{L} = \frac{1}{W} \frac{1}{H} \sum_{i=1}^{W}\sum_{j=1}^{H}(Y\textsubscript{i,j}- \widehat{Y}\textsubscript{i,j})^2    
\end{equation} 
where W and H are the width and height of a heat-map image, $Y$ is the ground-truth image, while $\widehat{Y}$ is the predicted output of STAR. 

\noindent\textit{Training and testing}: We trained two different models for hourly predictions and daily predictions. For hourly predictions, our model predicted 24 hours in one execution time. In long-term prediction, it generated output images corresponding to 24 hours of the predicted day. Accordingly, the predicted value of each day was averaged from outputs of corresponding hours on that day. Pixels of generated images were in the range of 0 to 1. After being aggregated into predicted values for districts, they were denormalized to get actual air quality levels.

\noindent\textit{Experimental Environment}: We trained models on a GPU-server with Nvidia 1080Ti GPU and used TensorFlow 1.14.

\subsubsection{Evaluation Metric} 
Mean absolute error (MAE) used to evaluate model performance is defined as follows:
\begin{equation}
MAE = \frac{\sum_{i} \mid y\textsubscript{i} - \widehat{y}\textsubscript{i} \mid}{n}
\end{equation}
where $y\textsubscript{i}$ and $\widehat{y}\textsubscript{i}$ are the ground-truth value and predicted value of a time-step $i$, and n is the total number of predictions.

\subsection{Experimental Results}
In this part, we present the experimental results of two datasets. In experimenting with the Seoul dataset, all prediction results were concentration values ($\mu g$/$m^3$). Whereas, we compared our model performance with FFA's method on Chinese AQI standards in China 1-year dataset. We used the following terms to indicate the difference in time: (1) short term: $\leq$ 8 hours (2) middle term: 9-24 hours (3) long term: $\geq$ 1 day. \\

Before moving to experimental results, we demonstrate the statistical information of PM\textsubscript{2.5} and PM\textsubscript{10} in Seoul (2014-2018) and Beijing (May 2014 - April 2015). In Seoul, we calculated the mean ($\mu$) and standard derivation ($\sigma$) values in particle concentrations ($\mu g/m^3$). In contrast, we computed these values in both particle concentrations and Chinese AQI standards for Beijing. As shown in \cref{statistic}, the mean value and standard derivation of PM\textsubscript{2.5} in Beijing are several times greater than in Seoul. We did not compute the statistical values for PM\textsubscript{10} in Beijing due to 38\% of missing data of this particulate matter. 

\setlength{\tabcolsep}{0.75em}
\begin{table}[t]
\centering
\caption{Statistic of PM\textsubscript{2.5} and PM\textsubscript{10} in Seoul (2014 - 2018) and Beijing (May 2014 - April 2015). } \label{statistic} 
\begin{tabular}{cccccc}
\hline
\multirow{2}{*}{City} & \multicolumn{2}{c}{PM\textsubscript{2.5}} & \multicolumn{2}{c}{PM\textsubscript{10}} & \multirow{2}{*}{Unit}  \\ \cline{2-5}
                      & $\mu$ & $\sigma$ & $\mu$ & $\sigma$ &          \\ 
\hline
Seoul                     & 24.75         & 16.54         & 45.99         & 34.07  & $\mu g$/$m^3$      \\ 
\hline
\multirow{2}{*}{Beijing}  & 83.12         & 80.12         & -         & - & $\mu g$/$m^3$        \\ 
                          & 106.67         & 91.79         & -         & - & China AQI         \\
\hline
\end{tabular}
\end{table}

\subsubsection{Comparing with Different Baselines} 
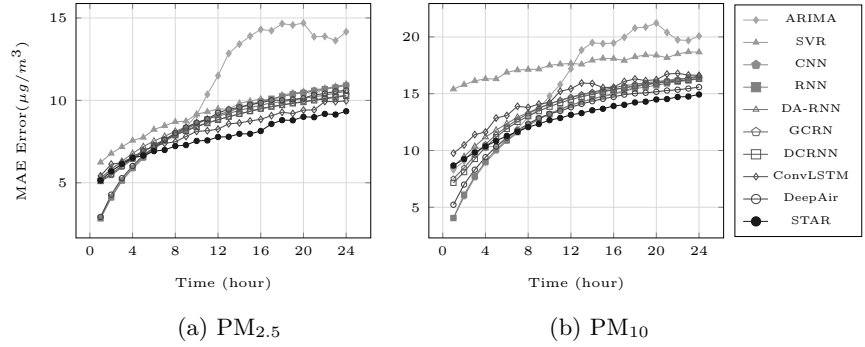
\begin{figure}[ht]
    \centering
    \begin{subfigure}[t]{0.492\linewidth}
    \begin{tikzpicture}
\begin{axis}[
    mark size=1pt,
    xtick={0,4,8,12,16,20,24},
    xlabel=Time (hour),ylabel=MAE Error($\mu g$/$m^3$)]
    \addplot[color=black!35!white,mark=diamond*, mark options={mark size=1.25pt}] coordinates{
        (1,5.28)
        (2,5.83)
        (3,6.24)
        (4,6.54)
        (5,6.62)
        (6,7.09)
        (7,7.58)
        (8,7.95)
        (9,8.36)
        (10,9.12)
        (11,10.36)
        (12,11.50)
        (13,12.85)
        (14,13.41)
        (15,13.90)
        (16,14.30)
        (17,14.23)
        (18,14.65)
        (19,14.57)
        (20,14.69)
        (21,13.86)
        (22,13.88)
        (23,13.63)
        (24,14.16)
    };
    \addlegendentry{ARIMA}
    \addplot[color=black!40!white,mark=triangle*, mark options={mark size=1.25pt}] coordinates{
        (1,6.24)
        (2,6.78)
        (3,7.18)
        (4,7.57)
        (5,7.75)
        (6,8.23)
        (7,8.46)
        (8,8.70)
        (9,8.74)
        (10,9.16)
        (11,9.31)
        (12,9.49)
        (13,9.46)
        (14,9.82)
        (15,9.96)
        (16,10.08)
        (17,9.99)
        (18,10.32)
        (19,10.44)
        (20,10.52)
        (21,10.39)
        (22,10.73)
        (23,10.84)
        (24,10.99)
    };
    \addlegendentry{SVR}
    \addplot[color=black!45!white,mark=pentagon*] coordinates {
        (1,2.85)
        (2,4.08)
        (3,5.12)
        (4,5.86)
        (5,6.49)
        (6,7.05)
        (7,7.50)
        (8,7.87)
        (9,8.25)
        (10,8.57)
        (11,8.89)
        (12,9.10)
        (13,9.32)
        (14,9.57)
        (15,9.78)
        (16,9.91)
        (17,10.07)
        (18,10.21)
        (19,10.40)
        (20,10.50)
        (21,10.64)
        (22,10.72)
        (23,10.82)
        (24,10.92)
    };
    \addlegendentry{CNN}
    \addplot[color=black!50!white,mark=square*] coordinates{
        (1,2.84)
        (2,4.12)
        (3,5.14)
        (4,5.90)
        (5,6.56)
        (6,7.09)
        (7,7.57)
        (8,7.94)
        (9,8.30)
        (10,8.57)
        (11,8.85)
        (12,9.13)
        (13,9.32)
        (14,9.54)
        (15,9.74)
        (16,9.92)
        (17,10.10)
        (18,10.24)
        (19,10.38)
        (20,10.46)
        (21,10.58)
        (22,10.68)
        (23,10.78)
        (24,10.90)
    };
    \addlegendentry{RNN}
    \addplot[color=black!55!white,mark=triangle, mark options={mark size=1.25pt}] coordinates{
        (1,5.27)
        (2,5.84)
        (3,6.31)
        (4,6.78)
        (5,7.20)
        (6,7.53)
        (7,7.82)
        (8,8.13)
        (9,8.38)
        (10,8.64)
        (11,8.87)
        (12,9.05)
        (13,9.21)
        (14,9.39)
        (15,9.56)
        (16,9.70)
        (17,9.86)
        (18,9.91)
        (19,10.02)
        (20,10.12)
        (21,10.22)
        (22,10.31)
        (23,10.45)
        (24,10.54)
    };
    \addlegendentry{DA-RNN}
    \addplot[color=black!60!white,mark=pentagon] coordinates{
        (1,5.10)
        (2,5.44)
        (3,5.95)
        (4,6.43)
        (5,6.88)
        (6,7.28)
        (7,7.63)
        (8,7.93)
        (9,8.19)
        (10,8.43)
        (11,8.64)
        (12,8.83)
        (13,9.00)
        (14,9.17)
        (15,9.33)
        (16,9.48)
        (17,9.61)
        (18,9.72)
        (19,9.83)
        (20,9.93)
        (21,10.03)
        (22,10.13)
        (23,10.22)
        (24,10.31)
    };
    \addlegendentry{GCRN}
    \addplot[color=black!65!white,mark=square] coordinates{
        (1,5.11)
        (2,5.53)
        (3,6.04)
        (4,6.48)
        (5,6.89)
        (6,7.25)
        (7,7.59)
        (8,7.89)
        (9,8.15)
        (10,8.39)
        (11,8.62)
        (12,8.82)
        (13,9.01)
        (14,9.18)
        (15,9.35)
        (16,9.50)
        (17,9.62)
        (18,9.72)
        (19,9.82)
        (20,9.90)
        (21,10.00)
        (22,10.09)
        (23,10.18)
        (24,10.27)
    };
    \addlegendentry{DCRNN}
    \addplot[color=black!70!white,mark=diamond, mark options={mark size=1.25pt}] coordinates{
        (1,5.42)
        (2,6.13)
        (3,6.22)
        (4,6.56)
        (5,6.94)
        (6,7.22)
        (7,7.39)
        (8,7.49)
        (9,7.81)
        (10,8.12)
        (11,8.16)
        (12,8.25)
        (13,8.58)
        (14,8.63)
        (15,8.75)
        (16,8.85)
        (17,9.07)
        (18,9.28)
        (19,9.20)
        (20,9.41)
        (21,9.42)
        (22,9.93)
        (23,9.92)
        (24,9.97)
    };
    \addlegendentry{ConvLSTM}
    \addplot[color=black!75!white,mark=o] coordinates{
        (1,2.93)
        (2,4.28)
        (3,5.28)
        (4,6.02)
        (5,6.67)
        (6,7.14)
        (7,7.61)
        (8,8.02)
        (9,8.35)
        (10,8.65)
        (11,8.86)
        (12,9.25)
        (13,9.45)
        (14,9.58)
        (15,9.73)
        (16,9.85)
        (17,10.01)
        (18,10.01)
        (19,10.06)
        (20,10.13)
        (21,10.34)
        (22,10.41)
        (23,10.55)
        (24,10.60)
    };
    \addlegendentry{DeepAir}
    \addplot[color=black!90!white,mark=otimes*] coordinates{
        (1,5.15)
        (2,5.70)
        (3,6.14)
        (4,6.51)
        (5,6.65)
        (6,6.92)
        (7,6.99)
        (8,7.23)
        (9,7.29)
        (10,7.54)
        (11,7.57)
        (12,7.78)
        (13,7.79)
        (14,7.97)
        (15,7.97)
        (16,8.14)
        (17,8.57)
        (18,8.80)
        (19,8.79)
        (20,9.00)
        (21,8.98)
        (22,9.18)
        (23,9.14)
        (24,9.34)
    };
    \addlegendentry{STAR}
    \legend{};
\end{axis}
\end{tikzpicture}
\caption{PM\textsubscript{2.5}}
\label{pm25_hourly}
    \end{subfigure}
    \hspace*{-1.3cm}
    \begin{subfigure}[t]{0.492\linewidth}
    \begin{tikzpicture}
\begin{axis}[
    mark size=1pt,
    xtick={0,4,8,12,16,20,24},
    xlabel=Time (hour), ylabel=\empty]
    \addplot[color=black!35!white,mark=diamond*, mark options={mark size=1.25pt}] coordinates{
        (1,8.30)
        (2,9.07)
        (3,10.17)
        (4,10.26)
        (5,10.41)
        (6,10.95)
        (7,11.91)
        (8,12.94)
        (9,13.67)
        (10,14.80)
        (11,15.81)
        (12,17.20)
        (13,18.84)
        (14,19.49)
        (15,19.41)
        (16,19.46)
        (17,19.97)
        (18,20.78)
        (19,20.87)
        (20,21.22)
        (21,20.39)
        (22,19.74)
        (23,19.68)
        (24,20.08)
    };
    \addlegendentry{ARIMA}
    \addplot[color=black!40!white,mark=triangle*, mark options={mark size=1.25pt}] coordinates{
        (1,15.40)
        (2,15.80)
        (3,16.13)
        (4,16.31)
        (5,16.30)
        (6,16.88)
        (7,17.08)
        (8,17.13)
        (9,17.16)
        (10,17.50)
        (11,17.60)
        (12,17.67)
        (13,17.59)
        (14,17.94)
        (15,18.11)
        (16,18.10)
        (17,17.92)
        (18,18.25)
        (19,18.41)
        (20,18.38)
        (21,18.17)
        (22,18.50)
        (23,18.68)
        (24,18.66)
    };
    \addlegendentry{SVR}
    \addplot[color=black!45!white,mark=pentagon*] coordinates {
        (1,4.02)
        (2,5.94)
        (3,7.60)
        (4,8.91)
        (5,9.96)
        (6,10.83)
        (7,11.56)
        (8,12.18)
        (9,12.76)
        (10,13.24)
        (11,13.66)
        (12,14.00)
        (13,14.30)
        (14,14.53)
        (15,14.80)
        (16,15.02)
        (17,15.26)
        (18,15.57)
        (19,15.82)
        (20,15.97)
        (21,16.16)
        (22,16.24)
        (23,16.36)
        (24,16.43)
    };
    \addlegendentry{CNN}
    \addplot[color=black!50!white,mark=square*] coordinates{
        (1,4.05)
        (2,6.11)
        (3,7.73)
        (4,9.01)
        (5,10.10)
        (6,10.90)
        (7,11.64)
        (8,12.23)
        (9,12.78)
        (10,13.19)
        (11,13.61)
        (12,14.00)
        (13,14.27)
        (14,14.55)
        (15,14.75)
        (16,14.95)
        (17,15.23)
        (18,15.45)
        (19,15.59)
        (20,15.74)
        (21,15.91)
        (22,16.06)
        (23,16.16)
        (24,16.32)
    };
    \addlegendentry{RNN}
    \addplot[color=black!55!white,mark=triangle, mark options={mark size=1.25pt}] coordinates{
        (1,8.59)
        (2,9.46)
        (3,10.36)
        (4,11.18)
        (5,11.79)
        (6,12.36)
        (7,12.93)
        (8,13.34)
        (9,13.75)
        (10,14.10)
        (11,14.41)
        (12,14.66)
        (13,14.89)
        (14,15.10)
        (15,15.23)
        (16,15.39)
        (17,15.51)
        (18,15.69)
        (19,15.84)
        (20,15.94)
        (21,16.09)
        (22,16.23)
        (23,16.40)
        (24,16.57)
    };
    \addlegendentry{DA-RNN}
    \addplot[color=black!60!white,mark=pentagon] coordinates{
        (1,7.52)
        (2,8.47)
        (3,9.57)
        (4,10.55)
        (5,11.39)
        (6,12.10)
        (7,12.72)
        (8,13.24)
        (9,13.71)
        (10,14.10)
        (11,14.43)
        (12,14.70)
        (13,14.94)
        (14,15.14)
        (15,15.34)
        (16,15.53)
        (17,15.70)
        (18,15.83)
        (19,15.96)
        (20,16.06)
        (21,16.17)
        (22,16.27)
        (23,16.36)
        (24,16.44)
    };
    \addlegendentry{GCRN}
    \addplot[color=black!65!white,mark=square] coordinates{
        (1,7.15)
        (2,8.10)
        (3,9.26)
        (4,10.28)
        (5,11.17)
        (6,11.91)
        (7,12.52)
        (8,13.03)
        (9,13.47)
        (10,13.86)
        (11,14.20)
        (12,14.48)
        (13,14.73)
        (14,14.95)
        (15,15.15)
        (16,15.35)
        (17,15.52)
        (18,15.68)
        (19,15.80)
        (20,15.91)
        (21,16.03)
        (22,16.12)
        (23,16.22)
        (24,16.32)
    };
    \addlegendentry{DCRNN}
    \addplot[color=black!70!white,mark=diamond, mark options={mark size=1.25pt}] coordinates{
        (1,9.76)
        (2,10.48)
        (3,11.40)
        (4,11.63)
        (5,12.86)
        (6,13.08)
        (7,13.89)
        (8,13.79)
        (9,14.09)
        (10,14.22)
        (11,15.21)
        (12,15.43)
        (13,15.94)
        (14,15.90)
        (15,15.54)
        (16,15.57)
        (17,16.08)
        (18,16.30)
        (19,16.16)
        (20,16.23)
        (21,16.73)
        (22,16.79)
        (23,16.65)
        (24,16.61)
    };
    \addlegendentry{ConvLSTM}
    \addplot[color=black!75!white,mark=o] coordinates{
        (1,5.22)
        (2,7.00)
        (3,8.31)
        (4,9.41)
        (5,10.34)
        (6,11.13)
        (7,11.77)
        (8,12.35)
        (9,12.85)
        (10,13.22)
        (11,13.48)
        (12,13.81)
        (13,14.10)
        (14,14.34)
        (15,14.55)
        (16,14.72)
        (17,14.92)
        (18,15.03)
        (19,15.08)
        (20,15.17)
        (21,15.28)
        (22,15.35)
        (23,15.45)
        (24,15.58)
    };
    \addlegendentry{DeepAir}
    \addplot[color=black!90!white,mark=otimes*] coordinates{
        (1,8.68)
        (2,9.25)
        (3,9.83)
        (4,10.37)
        (5,10.83)
        (6,11.29)
        (7,11.65)
        (8,12.06)
        (9,12.33)
        (10,12.67)
        (11,12.86)
        (12,13.14)
        (13,13.29)
        (14,13.54)
        (15,13.67)
        (16,13.92)
        (17,14.02)
        (18,14.23)
        (19,14.30)
        (20,14.49)
        (21,14.53)
        (22,14.71)
        (23,14.75)
        (24,14.93)
    };
    \addlegendentry{STAR}
\end{axis}
\end{tikzpicture}
\caption{PM\textsubscript{10}}
\label{pm10_hourly}
    \end{subfigure}
    \caption{Comparison of STAR with other methods on hourly predictions. Its performance is better than baselines after the 7\textsuperscript{th} hour.}
    \label{hourly_baseline_pred}
\end{figure}

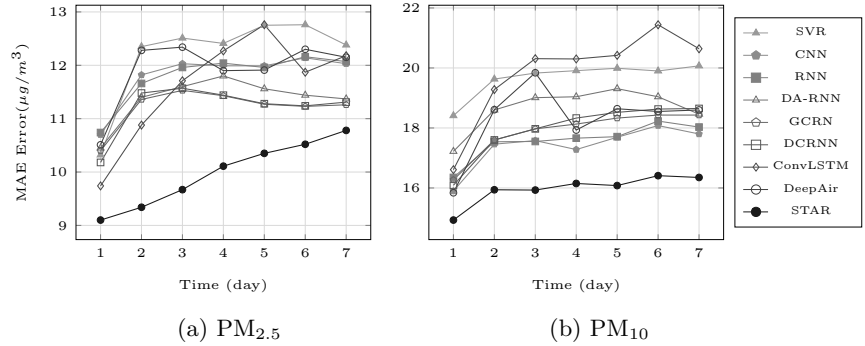
\begin{figure}[htbp]
    \centering
    \begin{subfigure}[t]{0.492\linewidth}
    \begin{tikzpicture}
\begin{axis}[
    mark size=1.25pt,
    xtick={1,2,3,4,5,6,7},
    ytick={8,9,10,11,12,13},
    ylabel=MAE Error($\mu g$/$m^3$),
    xlabel=Time (day),
    ]
    \addplot[color=black!40!white,mark=triangle*, mark options={mark size=1.5pt}] coordinates{
        (1,10.32)
        (2,12.35)
        (3,12.51)
        (4,12.41)
        (5,12.75)
        (6,12.76)
        (7,12.38)
    };
    \addlegendentry{SVR}
    \addplot[color=black!45!white,mark=pentagon*] coordinates {
        (1,10.70)
        (2,11.82)
        (3,12.03)
        (4,11.99)
        (5,11.99)
        (6,12.14)
        (7,12.03)
    };
    \addlegendentry{CNN}
    \addplot[color=black!50!white,mark=square*] coordinates{
        (1,10.74)
        (2,11.66)
        (3,11.96)
        (4,12.04)
        (5,11.96)
        (6,12.16)
        (7,12.07)
    };
    \addlegendentry{RNN}
    \addplot[color=black!55!white,mark=triangle, mark options={mark size=1.5pt}] coordinates{
        (1,10.45)
        (2,11.40)
        (3,11.60)
        (4,11.80)
        (5,11.56)
        (6,11.44)
        (7,11.37)
    };
    \addlegendentry{DA-RNN}
    \addplot[color=black!60!white,mark=pentagon] coordinates{
        (1,10.40)
        (2,11.36)
        (3,11.53)
        (4,11.43)
        (5,11.27)
        (6,11.23)
        (7,11.26)
    };
    \addlegendentry{GCRN}
    \addplot[color=black!65!white,mark=square] coordinates{
        (1,10.18)
        (2,11.48)
        (3,11.57)
        (4,11.44)
        (5,11.28)
        (6,11.24)
        (7,11.31)
    };
    \addlegendentry{DCRNN}
    \addplot[color=black!70!white,mark=diamond, mark options={mark size=1.5pt}] coordinates{
        (1, 9.74)
        (2, 10.88)
        (3, 11.71)
        (4, 12.27)
        (5, 12.76)
        (6, 11.87)
        (7, 12.18)
    };
    \addlegendentry{ConvLSTM}
    \addplot[color=black!75!white,mark=o] coordinates{
        (1,10.51)
        (2,12.28)
        (3,12.34)
        (4,11.90)
        (5,11.91)
        (6,12.30)
        (7,12.15)
    };
    \addlegendentry{DeepAir}
    \addplot[color=black!90!white,mark=otimes*] coordinates{
        (1,9.10)
        (2,9.34)
        (3,9.67)
        (4,10.11)
        (5,10.35)
        (6,10.52)
        (7,10.78)
    };
    \addlegendentry{STAR}
    \legend{};
\end{axis}
\end{tikzpicture}
\caption{PM\textsubscript{2.5}}
\label{pm25_daily}
    \end{subfigure}
    \hspace*{-1.3cm}
    \begin{subfigure}[t]{0.492\linewidth}
    \begin{tikzpicture}
\begin{axis}[
    mark size=1.25pt,
    xtick={1,2,3,4,5,6,7},
    xlabel=Time (day),
    ylabel=\empty]
    \addplot[color=black!40!white,mark=triangle*, mark options={mark size=1.5pt}] coordinates{
        (1,18.41)
        (2,19.63)
        (3,19.83)
        (4,19.91)
        (5,19.99)
        (6,19.90)
        (7,20.07)
    };
    \addlegendentry{SVR}
    \addplot[color=black!45!white,mark=pentagon*] coordinates {
        (1,15.90)
        (2,17.46)
        (3,17.58)
        (4,17.28)
        (5,17.69)
        (6,18.08)
        (7,17.80)
    };
    \addlegendentry{CNN}
    \addplot[color=black!50!white,mark=square*] coordinates{
        (1,16.35)
        (2,17.54)
        (3,17.55)
        (4,17.66)
        (5,17.71)
        (6,18.23)
        (7,18.03)   
    };
    \addlegendentry{RNN}
    \addplot[color=black!55!white,mark=triangle, mark options={mark size=1.5pt}] coordinates{
        (1,17.22)
        (2,18.60)
        (3,19.01)
        (4,19.04)
        (5,19.31)
        (6,19.04)
        (7,18.46)
    };
    \addlegendentry{DA-RNN}
    \addplot[color=black!60!white,mark=pentagon] coordinates{
        (1,16.27)
        (2,17.60)
        (3,17.97)
        (4,18.13)
        (5,18.33)
        (6,18.43)
        (7,18.43)
    };
    \addlegendentry{GCRN}
    \addplot[color=black!65!white,mark=square] coordinates{
        (1,16.09)
        (2,17.60)
        (3,17.97)
        (4,18.33)
        (5,18.52)
        (6,18.63)
        (7,18.65)
    };
    \addlegendentry{DCRNN}
    \addplot[color=black!70!white,mark=diamond, mark options={mark size=1.5pt}] coordinates{
        (1,16.61)
        (2,19.28)
        (3,20.31)
        (4,20.30)
        (5,20.42)
        (6,21.44)
        (7,20.64) 
    };
    \addlegendentry{ConvLSTM}
    \addplot[color=black!75!white,mark=o] coordinates{
        (1,15.84)
        (2,18.61)
        (3,19.84)
        (4,17.93)
        (5,18.64)
        (6,18.55)
        (7,18.60)
    };
    \addlegendentry{DeepAir}
    \addplot[color=black!90!white,mark=otimes*] coordinates{
        (1,14.93)
        (2,15.94)
        (3,15.93)
        (4,16.15)
        (5,16.08)
        (6,16.41)
        (7,16.35)
    };
    \addlegendentry{STAR}
\end{axis}
\end{tikzpicture}
\caption{PM\textsubscript{10}}
\label{pm10_daily}
    \end{subfigure}
    \caption{Comparison of STAR with other methods on daily predictions. Unlike the short-term prediction, STAR outperforms others by 11\% and 8.2\% for PM\textsubscript{2.5} and PM\textsubscript{10}, respectively.}
    \label{daily_baseline_pred}
\end{figure}

We compared our model with these methods in hourly prediction for up to twenty-four hours and long-term predictions for up to seven days ahead. In hourly prediction, our model performs better than other methods in the middle term, while baseline models, especially time-series methods, outperform STAR in the short term, as presented in \cref{hourly_baseline_pred}. In long-term prediction, STAR is superior to other methods, as shown in \cref{daily_baseline_pred}. To properly compare STAR with other methods, we averaged MAE scores of seven days for all methods. The proposed method achieves 11\% and 8.2\% improvement in MAE scores for PM\textsubscript{2.5} and PM\textsubscript{10}, respectively. To note that, we conducted multiple experiments to achieve the best results for baselines with appropriate settings.

We observed that PM\textsubscript{10} and PM\textsubscript{2.5} levels biased towards their levels in the latest hours, while their long-term levels significantly fluctuated due to the complex interactions of various factors, especially in the spatial dimension. Most of the methods focused on short-term prediction leading to better performance than STAR, while their results in the long-term prediction were unstable and ineffective. On the contrary, STAR focused on long-term variations of AQ since transforming monitoring data into heat-map images allowed our model to capture spatial relations efficiently, thus resulting in its superiority. However, CNN-generator tended to spatially average the predicted AQ levels of districts, while the dynamics of factors among areas were low in the short term. Therefore, the spatial learning process caused STAR to be less sensitive to the short-term variations, which affected the prediction accuracy. 

\begin{figure}[t!]
    \begin{subfigure}[t]{0.492\linewidth}
    \begin{tikzpicture}
\begin{axis}[
    mark size=1.25pt,
    legend style={draw=none,fill=white,at={(1.01,0.43)},anchor=north east,legend columns=1,nodes={scale=0.5, transform shape},column sep=0.08cm},
    xtick={0,4,8,12,16,20,24},
    xlabel=Time (hour),ylabel=MAE Error($\mu g$/$m^3$)]
    \addplot[color=black!35!white,mark=diamond*, mark options={mark size=1.5pt}] coordinates{
        (1,9.359033)
        (2,11.574743)
        (3,13.417356)
        (4,14.979836)
        (5,16.225544)
        (6,17.235910)
        (7,18.065700)
        (8,18.757858)
        (9,19.314370)
        (10,19.764057)
        (11,20.075762)
        (12,20.482265)
        (13,20.870178)
        (14,21.200603)
        (15,21.509810)
        (16,21.854765)
        (17,22.137222)
        (18,22.391014)
        (19,22.584402)
        (20,22.933638)
        (21,23.199001)
        (22,23.434944)
        (23,23.609552)
        (24,23.839533)
    };
    \addlegendentry{$\mathcal{I}$ + $\mathcal{M}_{all}$}
    \addplot[color=black!50!white,mark=pentagon*] coordinates {
        (1,8.309629)
        (2,10.723361)
        (3,12.669115)
        (4,14.187270)
        (5,15.455465)
        (6,16.485120)
        (7,17.268475)
        (8,17.850506)
        (9,18.308630)
        (10,18.641634)
        (11,18.947962)
        (12,19.225570)
        (13,19.469658)
        (14,19.717950)
        (15,19.903803)
        (16,20.096586)
        (17,20.272438)
        (18,20.436150)
        (19,20.593359)
        (20,20.812744)
        (21,20.988316)
        (22,21.180796)
        (23,21.336645)
        (24,21.524591)
    };
    \addlegendentry{$\mathcal{I}$ + $\mathcal{M}_{cloud}$}
    \addplot[color=black!65!white,mark=square*] coordinates{
        (1,7.657894)
        (2,9.576390)
        (3,11.179051)
        (4,12.455070)
        (5,13.557194)
        (6,14.530895)
        (7,15.312683)
        (8,15.948348)
        (9,16.420513)
        (10,16.825893)
        (11,17.106735)
        (12,17.390623)
        (13,17.628057)
        (14,17.884464)
        (15,18.132368)
        (16,18.327030)
        (17,18.559822)
        (18,18.800705)
        (19,19.065580)
        (20,19.281445)
        (21,19.451500)
        (22,19.562240)
        (23,19.690250)
        (24,19.813763)
    };
    \addlegendentry{$\mathcal{I}$ + $\mathcal{M}_{humid}$}
    \addplot[color=black!90!white,mark=otimes*] coordinates{
        (1,4.730146)
        (2,6.839430)
        (3,8.479495)
        (4,9.717571)
        (5,10.754258)
        (6,11.553611)
        (7,12.280178)
        (8,12.892530)
        (9,13.390781)
        (10,13.810783)
        (11,14.123075)
        (12,14.431396)
        (13,14.729898)
        (14,15.042385)
        (15,15.351067)
        (16,15.540572)
        (17,15.731931)
        (18,15.929674)
        (19,16.092257)
        (20,16.275585)
        (21,16.414486)
        (22,16.556381)
        (23,16.692130)
        (24,16.800283)
    };
    \addlegendentry{$\mathcal{I}$ }
    \legend{}
\end{axis}
\end{tikzpicture}
\caption{\footnotesize{RNN for PM\textsubscript{10}}}
\label{pm10_rnn}
    \end{subfigure}
    \hspace{-1.3cm}
    \begin{subfigure}[t]{0.492\linewidth}
    \begin{tikzpicture}
\begin{axis}[
    mark size=1.25pt,
    xtick={0,4,8,12,16,20,24},
    legend style={at={(1.54,0.5)}},
    xlabel=Time (hour),ylabel=\empty]
    \addplot[color=black!35!white,mark=diamond*, mark options={mark size=1.5}] coordinates{
       (1,7.010415)
        (2,8.312773)
        (3,9.522222)
        (4,10.472713)
        (5,11.421349)
        (6,12.123796)
        (7,12.818340)
        (8,13.384928)
        (9,13.928663)
        (10,14.401385)
        (11,14.822163)
        (12,15.266060)
        (13,15.638344)
        (14,15.972769)
        (15,16.394073)
        (16,16.710228)
        (17,16.931154)
        (18,17.267225)
        (19,17.508537)
        (20,17.778180)
        (21,18.043947)
        (22,18.313889)
        (23,18.559132)
        (24,18.733633)
    };
    \addlegendentry{$\mathcal{I}$ + $\mathcal{M}_{all}$}
    \addplot[color=black!50!white,mark=pentagon*] coordinates {
        (1,7.265350)
        (2,8.551701)
        (3,9.700214)
        (4,10.573441)
        (5,11.406486)
        (6,12.100245)
        (7,12.750420)
        (8,13.374098)
        (9,13.949236)
        (10,14.423423)
        (11,14.848371)
        (12,15.226833)
        (13,15.526502)
        (14,15.777521)
        (15,16.137062)
        (16,16.476730)
        (17,16.713186)
        (18,16.919872)
        (19,17.076338)
        (20,17.273836)
        (21,17.428875)
        (22,17.578620)
        (23,17.784527)
        (24,18.019672)
    };
    \addlegendentry{$\mathcal{I}$ + $\mathcal{M}_{cloud}$}
    \addplot[color=black!65!white,mark=square*] coordinates{
        (1,9.345508)
        (2,10.340780)
        (3,11.228347)
        (4,11.913343)
        (5,12.538794)
        (6,13.020899)
        (7,13.417246)
        (8,13.825867)
        (9,14.168839)
        (10,14.516127)
        (11,14.743214)
        (12,14.885002)
        (13,15.041653)
        (14,15.218235)
        (15,15.462066)
        (16,15.677273)
        (17,15.913074)
        (18,16.110230)
        (19,16.258606)
        (20,16.384893)
        (21,16.580332)
        (22,16.706587)
        (23,16.860638)
        (24,17.012209)
    };
    \addlegendentry{$\mathcal{I}$ + $\mathcal{M}_{humid}$}
    \addplot[color=black!90!white,mark=otimes*] coordinates{
        (1,5.084697)
        (2,6.899759)
        (3,8.318289)
        (4,9.355106)
        (5,10.276305)
        (6,11.080611)
        (7,11.752442)
        (8,12.323289)
        (9,12.856550)
        (10,13.273968)
        (11,13.499965)
        (12,13.769362)
        (13,14.057901)
        (14,14.259722)
        (15,14.557056)
        (16,14.830905)
        (17,15.067675)
        (18,15.223316)
        (19,15.356266)
        (20,15.520064)
        (21,15.680425)
        (22,15.826393)
        (23,15.932642)
        (24,16.042564)
    };
    \addlegendentry{$\mathcal{I}$}    
\end{axis}
\end{tikzpicture}
\caption{CNN for \footnotesize{PM\textsubscript{10}}}
\label{pm10_cnn}
    \end{subfigure}
    \caption{The performance of RNN and 1-D CNN on hourly predictions with meteorology data in input vectors. When adding meteorological features to these models, their prediction errors were increased. }
    \label{pm_baseline}
\end{figure}
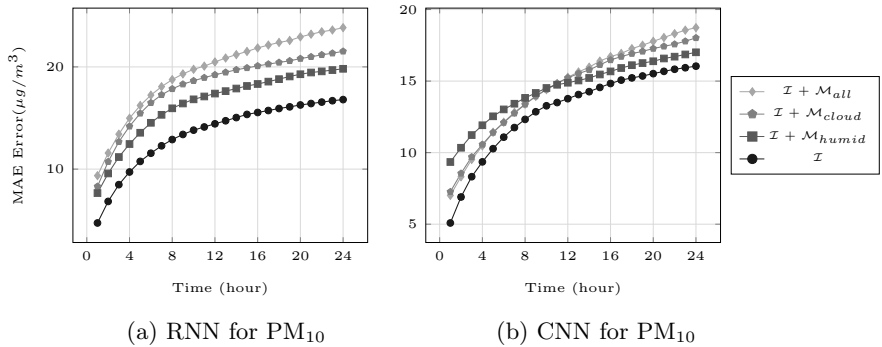

Practically, it is infeasible to make a model that can predict well future AQ levels in both the short-to-middle term and the long term. Furthermore, the bias-variance trade-off is a fundamental concern in the classical statistical learning theory, which describes the decline of performance as the model size increases. More predictors and more data do not guarantee better prediction results, as studied in \cite{nakkiran2019deep}. Therefore, baseline models such as ARIMA, SVR, CNN, and RNN were inefficient in middle-to-long-term predictions since they were not able to handle numerous predictors of different sources, thus cannot exploit valuable information from various factors. For instance, conventional time-series methods such as RNN or 1-D CNN worked ineffectively when meteorological information or the neighboring air pollution data was embedded in the model, as presented in \cref{pm_baseline}. Therefore, using only one model was not suitable for assessing the impact of critical factors on AQ prediction. 

STAR's results suggested that using heat-map images to model spatial correlation was an appropriate solution for long-term predictions. We directly compared its performance to ConvLSTM and DeepAir. The results of ConvLSTM in the first two days were better than other baselines, but the prediction error increased remarkably after that. The reason behind it is differences in the decoder structure of ConvLSTM and STAR as well as the order of the spatial information representation process. Despite the comparable performance in the short-to-middle-term prediction, the long-term prediction results of DeepAir's were also unstable. We guessed that the decline of performance resulted from the bias in predictions, which focused on the short-to-middle term. Besides, graph-based methods such as GCRN or DCRNN were not appropriate for the Seoul dataset since the geographical information was missing. 

\subsubsection{Experiments on China 1-year Dataset}

\begin{figure}[t!]
    \centering
    \begin{tikzpicture}
    \begin{axis}[
        width=6.9cm,
        height=4.5cm,
        mark size=1.5pt,
        legend style={at={(1.53,0.5)}},
        xlabel=Time (h),ylabel=MAE Error(China AQI),
        ytick={20,30,40,50,60,70},
        xtick={1,2,3,4},
        xticklabels={1-6,7-12,13-24,25-48}]
        \addplot[color=black!45!white,mark=pentagon] coordinates{
            (1, 30.12745833)
            (2,50.88106667)
            (3,66.16172)
            (4,73.45803333)
        };
        \addlegendentry{RNN}
        \addplot[color=black!55!white,mark=square] coordinates{
            (1, 35.41169)
            (2,53.89563833)
            (3,64.55161333)
            (4,70.99579667)
        };
        \addlegendentry{CNN}
        \addplot[color=black!60!white,mark=diamond*, mark options={mark size=2pt}] coordinates{
            (1,23.7)
            (2,52.4)
            (3,63.9)
            (4,69)
        };
        \addlegendentry{FFA \cite{zheng2015forecasting} }
        \addplot[color=black!75!white,mark=square*] coordinates{
            (1, 26.34385242)
            (2, 44.08434215)
            (3, 56.55746021)
            (4, 57.71045109)
        };
        \addlegendentry{STAR China Only}
        \addplot[color=black!90!white,mark=otimes*] coordinates{
            (1, 25.34490236)
            (2, 36.14975428)
            (3, 47.55584899)
            (4, 51.71998416)
        };
        \addlegendentry{STAR Seoul Transfer}
        \end{axis}
    \end{tikzpicture}
\caption{Comparison of our model with FFA and two baselines on PM\textsubscript{2.5} predictions in China 1-year dataset. Our model performance with transfer learning outperforms other methods.}
\label{china_ffa}
\end{figure}
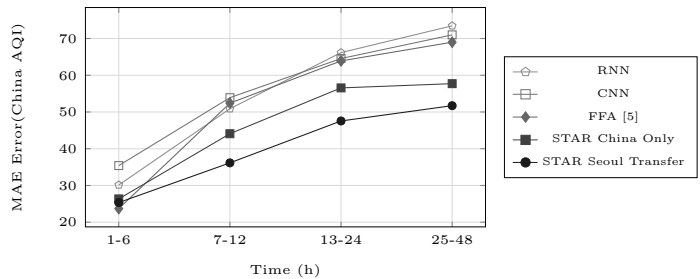
\pgfplotsset{width=5.3cm,height=4.6cm}

We conducted two experiments on this dataset to evaluate model robustness. First, we conducted an experiment, which trained the model from scratch, named as \textit{China Only}. In the China-only setting, we trained our model with 8 months and tested with 4 months as similar to the Zheng et al., 2015 (FFA). Secondly, we checked our model's capability of dealing with the lack of training data by applying transfer learning on Seoul dataset to China dataset, called \textit{Seoul Transfer}. In this experiment, we pre-trained the network on Seoul dataset for 20 epochs then re-trained it on China dataset for 300 epochs. Both settings outperformed the previous method, and the transfer learning setting is superior to others, as shown in \cref{china_ffa}. FFA is a simple method that cannot capture complex interactions among factors, especially in the spatial dimension. Besides, transfer learning is an appropriate approach to deal with the lack of data in AQ prediction. However, experimental results also emphasized our model weakness in short-term predictions.

\subsubsection{Assessing the Impact of Critical Factors} 
According to \cite{e88189}, the dispersion and formation of particulate matters depend on time, wind speed, precipitation, and humidity. Furthermore, PM\textsubscript{2.5} can travel from hundreds to thousands of kilometers and remain suspended in the air for weeks. Conversely, PM\textsubscript{10} can only disperse up to a few hundred kilometers and persist for a period ranging from a few minutes to several days. To assess the impact of critical factors on future AQ levels, we performed experiments on all possible combinations of data sources $\mathcal{I}$, $\mathcal{M}$, and $\mathcal{N}$. These combinations can be divided into two categories (1) involving $\mathcal{I}$ and (2) canceling $\mathcal{I}$ due to the similar trend of prediction errors. 

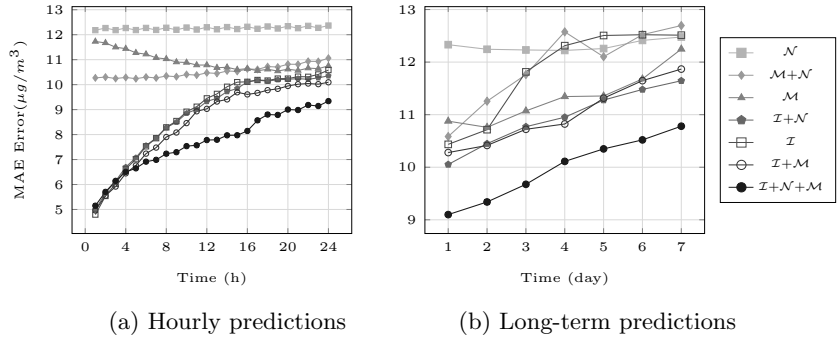
\begin{figure*}[ht]
    \centering
    \begin{subfigure}[t]{0.492\linewidth}
    \begin{tikzpicture}
\begin{axis}[
    mark size=1pt,
    ytick={5,6,7,8,9,10,11,12,13},
    xtick={0,4,8,12,16,20,24},
    xlabel=Time (h),ylabel=MAE Error($\mu g$/$m^3$)]
    
    \addplot[color=black!30!white,mark=square*] coordinates{
        (1,12.18153678)
        (2,12.26596493)
        (3,12.1824674)
        (4,12.2675564)
        (5,12.18488596)
        (6,12.27120299)
        (7,12.19004773)
        (8,12.27786676)
        (9,12.19763403)
        (10,12.28670715)
        (11,12.2072049)
        (12,12.29696463)
        (13,12.21616484)
        (14,12.30567569)
        (15,12.22552599)
        (16,12.31715535)
        (17,12.23755598)
        (18,12.32909103)
        (19,12.24901097)
        (20,12.34171195)
        (21,12.26180697)
        (22,12.35422225)
        (23,12.27401766)
        (24,12.36629994)
    };
    \addlegendentry{\scalebox{0.6}{$\scriptscriptstyle\mathcal{N}$}}

    \addplot[color=black!40!white,mark=diamond*, mark options={mark size=1.25pt}] coordinates {
        (1,10.2707086)
        (2,10.30238699)
        (3,10.24642424)
        (4,10.2882851)
        (5,10.23871264)
        (6,10.30260495)
        (7,10.26885441)
        (8,10.34589731)
        (9,10.31891347)
        (10,10.40664997)
        (11,10.38167767)
        (12,10.47370387)
        (13,10.44800336)
        (14,10.54715752)
        (15,10.52850875)
        (16,10.63131255)
        (17,10.60901639)
        (18,10.72286118)
        (19,10.70552564)
        (20,10.81794211)
        (21,10.81311099)
        (22,10.93509015)
        (23,10.92876704)
    (24,11.06081669)
    };
    \addlegendentry{\scalebox{0.6}{$\scriptscriptstyle\mathcal{M} + \scriptscriptstyle\mathcal{N}$}}
    \addplot[color=black!50!white,mark=triangle*, mark options={mark size=1.25pt}] coordinates{
        (1,11.72793983)
        (2,11.6717708)
        (3,11.49899516)
        (4,11.44157458)
        (5,11.27421933)
        (6,11.21948428)
        (7,11.0743126)
        (8,11.03061953)
        (9,10.9041296)
        (10,10.88668089)
        (11,10.78440783)
        (12,10.78500341)
        (13,10.67701331)
        (14,10.69528228)
        (15,10.60154578)
        (16,10.64038635)
        (17,10.55644131)
        (18,10.61522071)
        (19,10.54574442)
        (20,10.61708015)
        (21,10.56630181)
        (22,10.66166687)
        (23,10.64118622)
        (24,10.75011104)
    };
    \addlegendentry{\scalebox{0.6}{$\scriptscriptstyle\mathcal{M}$}}
    
    \addplot[color=black!60!white,mark=pentagon*] coordinates{
        (1,4.99812966)
        (2,5.65187993)
        (3,6.13775544)
        (4,6.69849639)
        (5,7.07504659)
        (6,7.56202179)
        (7,7.86214104)
        (8,8.2777644)
        (9,8.49591172)
        (10,8.85282811)
        (11,9.01078112)
        (12,9.32585641)
        (13,9.45420573)
        (14,9.72179379)
        (15,9.84729295)
        (16,10.11077302)
        (17,10.17754023)
        (18,10.14100393)
        (19,10.197941)
        (20,10.21320968)
        (21,10.24114294)
        (22,10.21333165)
        (23,10.27271421)
    (24,10.35153439)
    };
    \addlegendentry{\scalebox{0.6}{$\scriptscriptstyle\mathcal{I} + \scriptscriptstyle\mathcal{N}$}}

    \addplot[color=black!70!white,mark=square] coordinates{
        (1,4.79423177)
        (2,5.54077216)
        (3,6.02762156)
        (4,6.62309857)
        (5,7.01568267)
        (6,7.52861416)
        (7,7.8546593)
        (8,8.28938525)
        (9,8.5425243)
        (10,8.91644786)
        (11,9.11586258)
        (12,9.45074344)
        (13,9.62869869)
        (14,9.91545266)
        (15,10.09933147)
        (16,10.11338377)
        (17,10.18574381)
        (18,10.16405055)
        (19,10.24259552)
        (20,10.24114294)
        (21,10.31970817)
        (22,10.3132622)
        (23,10.40849457)
    (24,10.58383494)
    };
    \addlegendentry{\scalebox{0.6}{$\scriptscriptstyle\mathcal{I}$}}

    \addplot[color=black!80!white,mark=o] coordinates{
        (1,4.9603181)
        (2,5.55078713)
        (3,5.91729183)
        (4,6.44716342)
        (5,6.76288387)
        (6,7.23689109)
        (7,7.47819779)
        (8,7.90128764)
        (9,8.08261057)
        (10,8.45904232)
        (11,8.93211289)
        (12,9.03046391)
        (13,9.32354835)
        (14,9.40890005)
        (15,9.69338443)
        (16,9.60906506)
        (17,9.66708256)
        (18,9.77879909)
        (19,9.83161812)
        (20,9.94705975)
        (21,10.01782515)
        (22,10.04433199)
        (23,10.01920317)
    (24,10.09441108)
    };
    \addlegendentry{\scalebox{0.6}{$\scriptscriptstyle\mathcal{I} + \scriptscriptstyle\mathcal{M}$}}

    \addplot[color=black!90!white,mark=otimes*] coordinates{
        (1,5.15086836)
        (2,5.70485351)
        (3,6.14485992)
        (4,6.506154943)
        (5,6.647860693)
        (6,6.915405693)
        (7,6.987812393)
        (8,7.234025207)
        (9,7.29439403)
        (10,7.536981293)
        (11,7.571520137)
        (12,7.780341617)
        (13,7.7921008)
        (14,7.970255473)
        (15,7.974045577)
        (16,8.143944197)
        (17,8.57082577)
        (18,8.80270606)
        (19,8.78655883)
        (20,9.00231971)
        (21,8.98355483)
        (22,9.1837943)
        (23,9.14391934)
    (24,9.34041734)
    };
    \addlegendentry{\scalebox{0.6}{$\scriptscriptstyle\mathcal{I} + \scriptscriptstyle\mathcal{N} + \scriptscriptstyle\mathcal{M}$}}

    \legend{};
    \end{axis}
\end{tikzpicture}
\caption{Hourly predictions}
\label{pm25_hourly_data}
    \end{subfigure}
    \hspace*{-1.3cm}
    \begin{subfigure}[t]{0.492\linewidth}
    \begin{tikzpicture}
\begin{axis}[
    mark size=1.25pt,
    xtick={1,2,3,4,5,6,7},
    ytick={4,5,6,7,8,9,10,11,12,13},
    xlabel=Time (day),ylabel=\empty]
    
    \addplot[color=black!30!white,mark=square*] coordinates{
        (1,12.3315133)
        (2,12.24406994)
        (3,12.23089911)
        (4,12.22217072)
        (5,12.25747732)
        (6,12.41295648)
        (7,12.4771038)
    };
    \addlegendentry{$\scriptscriptstyle\mathcal{N}$}

    \addplot[color=black!40!white,mark=diamond*, mark options={mark size=1.5pt}] coordinates {
        (1,10.5843214)
        (2,11.25579135)
        (3,11.76277758)
        (4,12.57520399)
        (5,12.10191821)
        (6,12.52189515)
        (7,12.69524395)
    };
    \addlegendentry{$\scriptscriptstyle\mathcal{M} + \scriptscriptstyle\mathcal{N}$}
    
    \addplot[color=black!50!white,mark=triangle*, mark options={mark size=1.5pt}] coordinates{
        (1,10.8768754)
        (2,10.75854641)
        (3,11.07155409)
        (4,11.34420414)
        (5,11.35541172)
        (6,11.67928162)
        (7,12.24744015)
    };
    \addlegendentry{$\scriptscriptstyle\mathcal{M}$}

    \addplot[color=black!60!white,mark=pentagon*] coordinates{
        (1,10.0526487)
        (2,10.44839437)
        (3,10.77129187)
        (4,10.95131986)
        (5,11.277642)
        (6,11.4793282)
        (7,11.64509724)
    };
    \addlegendentry{$\scriptscriptstyle\mathcal{I} + \scriptscriptstyle\mathcal{N}$}

    \addplot[color=black!70!white,mark=square] coordinates{
        (1,10.4351972)
        (2,10.7145084)
        (3,11.81390201)
        (4,12.31217338)
        (5,12.50623191)
        (6,12.52082708)
        (7,12.51232909)
    };
    \addlegendentry{$\scriptscriptstyle\mathcal{I}$}

    \addplot[color=black!80!white,mark=o] coordinates{
        (1,10.2791934)
        (2,10.4146659)
        (3,10.72195562)
        (4,10.82268891)
        (5,11.3137171)
        (6,11.64697714)
        (7,11.86723282)
    };
    \addlegendentry{$\scriptscriptstyle\mathcal{I} + \scriptscriptstyle\mathcal{M}$}

    \addplot[color=black!90!white,mark=otimes*] coordinates{
        (1,9.098752783)
        (2,9.339159773)
        (3,9.674939479)
        (4,10.11121074)
        (5,10.34891283)
        (6,10.52011834)
        (7,10.78025712)
    };
    \addlegendentry{$\scriptscriptstyle\mathcal{I} + \scriptscriptstyle\mathcal{N} + \scriptscriptstyle\mathcal{M}$}
    \end{axis}
\end{tikzpicture}
\caption{Long-term predictions}
\label{pm25_daily_data}
    \end{subfigure}
    \caption{An assessment of critical factors affecting short-to-long-term predictions of PM\textsubscript{2.5} levels in the Seoul dataset. It was more responsive to changes in the pattern of dominant factors than PM\textsubscript{10}. }
    \label{pm_data}
\end{figure*}

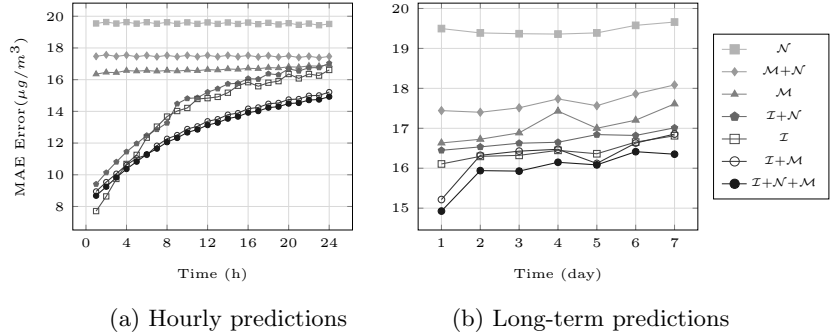
\begin{figure*}[hbtp]
    \centering
    \begin{subfigure}[t]{0.492\linewidth}
    \begin{tikzpicture}
\begin{axis}[
    mark size=1pt,
    ytick={8,10,12,14,16,18,20},
    xtick={0,4,8,12,16,20,24},
    xlabel=Time (h),
    ylabel=MAE Error($\mu g$/$m^3$)
    ]
    \addplot[color=black!30!white,mark=square*] coordinates{
        (1,19.54543636)
        (2,19.63558229)
        (3,19.54022982)
        (4,19.63014208)
        (5,19.53484697)
        (6,19.62467123)
        (7,19.53023771)
        (8,19.62012158)
        (9,19.52505823)
        (10,19.61441431)
        (11,19.52070109)
        (12,19.61115471)
        (13,19.5187416)
        (14,19.60858968)
        (15,19.51445874)
        (16,19.60290399)
        (17,19.50229906)
        (18,19.58581736)
        (19,19.483023)
        (20,19.56440035)
        (21,19.46023508)
        (22,19.53984871)
        (23,19.43518765)
        (24,19.51427094)
    };
    \addlegendentry{$\scriptscriptstyle\mathcal{N}$}
    
    \addplot[color=black!40!white,mark=diamond*, mark options={mark size=1.25}] coordinates {
        (1,17.473339)
        (2,17.56553467)
        (3,17.46975232)
        (4,17.5484919)
        (5,17.45102491)
        (6,17.53323868)
        (7,17.44492493)
        (8,17.53164824)
        (9,17.44645295)
        (10,17.5363827)
        (11,17.44836685)
        (12,17.5357321)
        (13,17.44687393)
        (14,17.53688887)
        (15,17.44626546)
        (16,17.52942477)
        (17,17.43308623)
        (18,17.51519231)
        (19,17.42267395)
        (20,17.49787715)
        (21,17.402531)
        (22,17.4792398)
        (23,17.3826587)
        (24,17.466157)
    };
    \addlegendentry{$\scriptscriptstyle\mathcal{M} + \scriptscriptstyle\mathcal{N}$}
    
    \addplot[color=black!50!white,mark=triangle*, mark options={mark size=1.25}] coordinates{
        (1,16.34979898)
        (2,16.46011089)
        (3,16.44650133)
        (4,16.57402576)
        (5,16.53906857)
        (6,16.57515281)
        (7,16.53143356)
        (8,16.57923654)
        (9,16.5408962)
        (10,16.59944889)
        (11,16.56783483)
        (12,16.6353011)
        (13,16.60467788)
        (14,16.67447198)
        (15,16.64806404)
        (16,16.71928012)
        (17,16.69201256)
        (18,16.76209898)
        (19,16.73736863)
        (20,16.81027173)
        (21,16.78790237)
        (22,16.86791002)
        (23,16.85315789)
        (24,16.93830612)
    };
    \addlegendentry{$\scriptscriptstyle\mathcal{M}$}
    
    \addplot[color=black!60!white,mark=pentagon*] coordinates{
        (1,9.40977183)
        (2,10.13582596)
        (3,10.80986307)
        (4,11.45001204)
        (5,11.96571715)
        (6,12.47985668)
        (7,12.86092563)
        (8,13.26884027)
        (9,14.47907786)
        (10,14.81263545)
        (11,14.85801271)
        (12,15.21230333)
        (13,15.42790104)
        (14,15.7332917)
        (15,15.77411405)
        (16,16.07002073)
        (17,16.03331531)
        (18,16.37266019)
        (19,16.30612862)
        (20,16.68359986)
        (21,16.54777131)
        (22,16.74948976)
        (23,16.76242295)
        (24,17.03490585)
    };
    \addlegendentry{$\scriptscriptstyle\mathcal{I} + \scriptscriptstyle\mathcal{N}$}
    
    \addplot[color=black!70!white,mark=square] coordinates{
        (1,7.70901529)
        (2,8.63663567)
        (3,9.73785967)
        (4,10.68446181)
        (5,11.255938967)
        (6,12.33852465)
        (7,13.03748863)
        (8,13.67144799)
        (9,14.02107786)
        (10,14.21463545)
        (11,14.78201271)
        (12,14.83230333)
        (13,14.91390104)
        (14,15.1762917)
        (15,15.61811405)
        (16,15.84102073)
        (17,15.58431531)
        (18,15.80766019)
        (19,15.90612862)
        (20,16.35659986)
        (21,16.07877131)
        (22,16.35248976)
        (23,16.25642295)
        (24,16.61290585)
    };
    \addlegendentry{$\scriptscriptstyle\mathcal{I}$}
    
    \addplot[color=black!80!white,mark=o] coordinates{
        (1,8.9489827)
        (2,9.53934862)
        (3,10.0672468)
        (4,10.62077631)
        (5,11.03690471)
        (6,11.252181336)
        (7,11.83284108)
        (8,12.27122307)
        (9,12.50535072)
        (10,12.88700142)
        (11,13.05630673)
        (12,13.37615588)
        (13,13.50625222)
        (14,13.79064535)
        (15,13.88924145)
        (16,14.17013593)
        (17,14.24615216)
        (18,14.48054251)
        (19,14.53064659)
        (20,14.74914761)
        (21,14.77677472)
        (22,14.98509351)
        (23,15.00890247)
        (24,15.2158817)
    };
    \addlegendentry{$\scriptscriptstyle\mathcal{I} + \scriptscriptstyle\mathcal{M}$}
    
    \addplot[color=black!90!white,mark=otimes*] coordinates{
        (1,8.67737166)
        (2,9.24617519)
        (3,9.8332539)
        (4,10.3650204)
        (5,10.82802096)
        (6,11.29210134)
        (7,11.64932176)
        (8,12.05893363)
        (9,12.32507786)
        (10,12.67063545)
        (11,12.85601271)
        (12,13.13530333)
        (13,13.28790104)
        (14,13.5432917)
        (15,13.66911405)
        (16,13.92302073)
        (17,14.02031531)
        (18,14.23466019)
        (19,14.29812862)
        (20,14.49259986)
        (21,14.53377131)
        (22,14.71348976)
        (23,14.74642295)
        (24,14.92690585)
    };
    \addlegendentry{$\scriptscriptstyle\mathcal{I} + \scriptscriptstyle\mathcal{N} + \scriptscriptstyle\mathcal{M}$}
    \legend{}
    \end{axis}
\end{tikzpicture}
\caption{Hourly predictions}
\label{pm10_hourly_data}
    \end{subfigure}
    \hspace*{-1.4cm}
    \begin{subfigure}[t]{0.492\linewidth}
    \begin{tikzpicture}
\begin{axis}[
    mark size=1.25pt,
    ytick={14,15,16,17,18,19,20},
    xtick={1,2,3,4,5,6,7},
    xlabel=Time (day),
    ylabel=\empty,
    ]
    
    \addplot[color=black!30!white,mark=square*] coordinates{
        (1,19.4964358)
        (2,19.38785105)
        (3,19.36781209)
        (4,19.35886765)
        (5,19.38843895)
        (6,19.57695931)
        (7,19.66015457)
    };
    \addlegendentry{$\scriptscriptstyle\mathcal{N}$}
    \addplot[color=black!40!white,mark=diamond*, mark options={mark size=1.5pt}] coordinates {
        (1,17.4426852)
        (2,17.40130215)
        (3,17.51305894)
        (4,17.73549386)
        (5,17.56488683)
        (6,17.85916236)
        (7,18.0861758)
    };
    \addlegendentry{$\scriptscriptstyle\mathcal{M} + \scriptscriptstyle\mathcal{N}$}
    \addplot[color=black!50!white,mark=triangle*, mark options={mark size=1.5pt}] coordinates{
        (1,16.63287068)
        (2,16.72559565)
        (3,16.88534873)
        (4,17.42985862)
        (5,17.00026354)
        (6,17.20354901)
        (7,17.6110411)
    };
    \addlegendentry{$\scriptscriptstyle\mathcal{M}$}
    \addplot[color=black!60!white,mark=pentagon*] coordinates{
    
    (1,16.4489395)
        (2,16.53460774)
        (3,16.6263855)
        (4,16.64993765)
        (5,16.84215014)
        (6,16.82082678)
        (7,17.0113745)
    };
    \addlegendentry{$\scriptscriptstyle\mathcal{I} + \scriptscriptstyle\mathcal{N}$}
    \addplot[color=black!70!white,mark=square] coordinates{
        (1,16.10727285)
        (2,16.29698411)
        (3,16.32234936)
        (4,16.4481656)
        (5,16.36244625)
        (6,16.65445068)
        (7,16.8127642)
    };
    \addlegendentry{$\scriptscriptstyle\mathcal{I}$}
    \addplot[color=black!80!white,mark=o] coordinates{
        (1,15.2158817)
        (2,16.32336996)
        (3,16.4276983)
        (4,16.47457217)
        (5,16.1200072)
        (6,16.63074777)
        (7,16.8545280)
    };
    \addlegendentry{$\scriptscriptstyle\mathcal{I} + \scriptscriptstyle\mathcal{M}$}
    \addplot[color=black!90!white,mark=otimes*] coordinates{
        (1,14.92690585)
        (2,15.94036996)
        (3,15.9276983)
        (4,16.14877217)
        (5,16.08385061)
        (6,16.41437553)
        (7,16.3514242)
    };
    \addlegendentry{$\scriptscriptstyle\mathcal{I} + \scriptscriptstyle\mathcal{N} + \scriptscriptstyle\mathcal{M}$}
    \end{axis}
\end{tikzpicture}
\caption{Long-term predictions}
\label{pm10_daily_data}
    \end{subfigure}
    \caption{An evaluation of the effects of critical factors on short-to-long-term predictions of PM\textsubscript{10} levels in the Seoul dataset. Similar to PM\textsubscript{2.5}, these effects on PM\textsubscript{10} are more discernible in the long term than the short-to-middle term.}
    \label{pm10_data}
\end{figure*}

\noindent\textit{Short-to-middle-term impact}. As shown in \cref{pm25_hourly_data,pm10_hourly_data}, local AQ data ($\mathcal{I}$), when included in input vectors (E.g., $\mathcal{I} + \mathcal{M}$, $\mathcal{I} + \mathcal{N}$), allowed the model to achieve superior performance compared to the other settings. We can infer that short-to-middle-term AQ variations firmly depended on the local AQ data.

As depicted in \cref{pm25_hourly_data}, the performance of PM\textsubscript{2.5} prediction using only meteorology data ($\mathcal{M}$) is defective in the short term, while it \textbf{}gradually improves in the middle term. These observations suggested the increased influence of meteorology on PM\textsubscript{2.5} predictions in the middle term. The parallel performance of experimental settings with $\mathcal{I}$, $\mathcal{I} + \mathcal{M}$, and $\mathcal{I} + \mathcal{N}$ indicated the corresponding impact of both local meteorological conditions and neighboring air pollution sources on short-to-middle-term PM\textsubscript{2.5} levels. Additionally, the outstanding performance of the experiment with all data sources emphasized the strong correlation of PM\textsubscript{2.5} variations with critical factors. Besides, the weak results of the setting using only $\mathcal{N}$ data showed that external air pollution sources had less influence on short-to-middle-term PM\textsubscript{2.5} levels than meteorological conditions. Therefore, short-to-middle-term PM\textsubscript{2.5} levels were the results of a combination of various factors.

Similarly, PM\textsubscript{10} levels were also connected with dominant influences in the short and middle term. However, the impact of each factor on short-to-middle-term PM\textsubscript{10} levels was not as discernible as PM\textsubscript{2.5}. As \cref{pm10_hourly_data} shows, the errors of experimental settings $\mathcal{N}$, $\mathcal{M}$, and $\mathcal{M + N}$ are constant and significant. Besides, the similar performance of two settings $\mathcal{I + N}$ and $\mathcal{I}$ emphasizes the negligible impact of external air pollution sources on the variations of short-to-middle-term PM\textsubscript{10}. Finally, the analogy of the performance of two settings $\mathcal{I + N + M}$ and $\mathcal{I + M}$ showed the strong correlation of meteorology and local observational AQI with PM\textsubscript{10} in the short and middle term.

\noindent\textit{Long-term impact}. \cref{pm25_daily_data,pm10_daily_data} showed apparent effects of dominant factors on long-term AQ variations. 

As depicted in \cref{pm25_daily_data}, meteorology and the external air pollution sources significantly impact the variations of long-term PM\textsubscript{2.5} levels. First, all experimental settings without the advent of $\mathcal{I}$ have similar performance with the prediction setting using only $\mathcal{I}$ data. Next, the comparable performance of $\mathcal{M} + \mathcal{N}$, $\mathcal{I} + \mathcal{M}$, and $\mathcal{I} + \mathcal{N}$ indicated the relative effects of critical factors on long-term PM\textsubscript{2.5} levels. Finally, similar to short-to-middle-term predictions, the distinctive results of the combination of all data sources emphasized the strong interconnectedness of meteorology, the external air pollution sources, and the future PM\textsubscript{2.5} levels.

Unlike long-term PM\textsubscript{2.5} predictions, the impact of critical factors on PM\textsubscript{10} levels was not evident, especially the external air pollution sources. From \cref{pm10_daily_data}, the experiment with only $\mathcal{N}$ provides the worst performance followed by the settings $\mathcal{M} + \mathcal{N}$ and $\mathcal{M}$. Additionally, the performance of the predictive model using only local AQ data ($\mathcal{I}$) is a bit worse than the results of the experiment with $\mathcal{I}$ and $\mathcal{N}$. These results showed the slight influence of the external air pollution sources on PM\textsubscript{10} levels as similar to short-to-middle-term predictions. Furthermore, the analogy of results of the two settings $\mathcal{I + N + M}$ and $\mathcal{I + M}$ indicated the direct influence of meteorology on the variations of long-term PM\textsubscript{10}. Finally, the distinguished performance of all data sources combination confirmed the accumulative impacts of all factors on PM\textsubscript{10} levels.

\section{Conclusion} \label{conclusion_part}
In this paper, we proposed a multimodal approach using various data sources for predicting short-to-long-term AQ levels. Unlike previous methods, our model transforms observational data into heat-map images to efficiently capture spatial relations. The experimental results showed that our model outperforms baselines and state-of-the-art methods, especially in long term predictions. The results on China 1-year dataset demonstrated the model robustness to similar datasets. These results open the door for dealing with the lack of data in AQ predictions for many urban areas. Next, we assessed the impact of dominant factors on AQ variations. Besides that, we realized that LSTM-based models are biased toward previous values when making predictions. However, future air quality does not entirely depend on previous time-steps due to the strong interconnectedness of AQ levels and many factors. Therefore, further studies are necessary to identify sudden changes in AQ levels along with short-to-long-term predictions.

\section*{Acknowledgments}

This work was supported by the New  Industry Promotion Program (1415158216, Development of Front/Side Camera Sensor for Autonomous Vehicle) funded by the Ministry of Trade, Industry \& Energy (MOTIE, Korea).

\bibliography{main}
\bibliographystyle{ieeetr}

\end{document}